\documentclass[final,5p,times,twocolumn]{elsarticle}
\usepackage{amssymb}
\usepackage{hyperref}
\usepackage{array}
\usepackage{multirow}
\usepackage{booktabs}
\usepackage{adjustbox}
\usepackage{soul}
\usepackage{color}
\usepackage{diagbox}
\usepackage{enumitem}
\usepackage[absolute]{textpos}

\textblockorigin{10mm}{10mm}
\journal{Signal Processing}

\begin{document}
\begin{textblock}{14}(4.5,6.5)
\textbf{Accepted to Appear in Signal Processing (Elsevier), September 2023}
\\ \small DOI: \url{https://doi.org/10.1016/j.sigpro.2023.109248}
\end{textblock}

\begin{frontmatter}

\title{Hyperspectral Image Denoising via Self-Modulating Convolutional Neural Networks}

\author[inst1,inst2]{Orhan Torun}
\author[inst2]{Seniha Esen Yuksel}
\author[inst3,inst6]{Erkut Erdem}
\author[inst4]{Nevrez Imamoglu}
\author[inst5,inst6]{Aykut Erdem}

\affiliation[inst1]{organization={Hacettepe University, Institute of Science},
            addressline={Beytepe}, 
            city={Ankara},
            postcode={06800}, 
            country={Turkey}}

\affiliation[inst2]{organization={Hacettepe University, Department of Electrical and Electronics Engineering},
            addressline={Beytepe}, 
            city={Ankara},
            postcode={06800}, 
            country={Turkey}}
\affiliation[inst3]{organization={Hacettepe University, Department of Computer Engineering},
            addressline={Beytepe}, 
            city={Ankara},
            postcode={06800}, 
            country={Turkey}}
\affiliation[inst4]{organization={National Institute of Advanced Industrial Science and Technology, Digital Architecture Research Center},
            city={Tokyo},
            postcode={135-0064}, 
            country={Japan}}
\affiliation[inst5]{organization={ Koç University,  Department of Computer Engineering},
            addressline={Sarıyer}, 
            city={Istanbul},
            postcode={34450}, 
            country={Turkey}}
\affiliation[inst6]{organization={ Koç University, Is Bank AI Center},
            addressline={Sarıyer}, 
            city={Istanbul},
            postcode={34450}, 
            country={Turkey}}

\begin{abstract}
\textcolor{black}{Compared to natural images, hyperspectral images (HSIs) consist of a large number of bands, with each band capturing different spectral information from a certain wavelength, even some beyond the visible spectrum. These characteristics of HSIs make them highly effective for remote sensing applications. That said, the existing hyperspectral imaging devices introduce severe degradation in HSIs. Hence, hyperspectral image denoising  has attracted lots of attention by the community lately. \textcolor{black}{While recent deep HSI denoising methods have provided effective solutions, their performance under real-life complex noise remains suboptimal, as they lack adaptability to new data.} To overcome these limitations, in our work, we introduce a self-modulating convolutional neural network which we refer to as SM-CNN, which utilizes correlated spectral and spatial information. At the core of the model lies a novel block, which we call spectral self-modulating residual block (SSMRB), that allows the network to transform the features in an adaptive manner based on the adjacent spectral data, enhancing the network's ability to handle complex noise. In particular, the introduction of SSMRB transforms our denoising network into a dynamic network that adapts its predicted features while denoising every input HSI with respect to its spatio-spectral characteristics. Experimental analysis on both synthetic and real data shows that the proposed SM-CNN outperforms other state-of-the-art HSI denoising methods both quantitatively and qualitatively on public benchmark datasets. Our code will be available at \url{https://github.com/orhan-t/SM-CNN}.}\\

\end{abstract}


\end{frontmatter}

\section{Introduction}\label{sec:1}

Hyperspectral images (HSIs) contain rich spectral and spatial information of a scene, which makes them useful for many applications \cite{kucuk2021total, qi2020deep, lu2023ensemble}. In remote sensing, HSI sensors are generally mounted on aircrafts, drones or satellites that present harsh operating environments for the sensors. Hence, during the acquisition process, HSI can be easily contaminated by noise that is caused by several different factors such as atmospheric absorption, temperature, lighting condition and sensor malfunctions. These environmental conditions and sensory malfunctions can cause various types of noise: (i) Gaussian noise (GN), (ii) stripe noise (SN), (iii)  dead pixel noise (DN), (iv) impulse noise (IN), and (v) the mixture of all these noise types \cite{rasti2018noise,zhang2019Hybrid}. Further, each band in HSI might be subject to different types and levels of noise \cite{bahraini2022bayesian}. As a result, noisy data often affects the performance of image interpretation and subsequent applications; thus noise reduction in HSI is considered an essential prepossessing step.

Recently, with the popularity of deep learning (DL), CNN based approaches have created a fresh wave of HSI denoising methods, demonstrating significant improvements over the traditional methods. These data-driven models automatically learn a mapping from noisy HSIs to clean HSIs. In order to make the network learn about this non-linear mapping, training pairs consisting of a large number of noisy and clean data are needed. However, collecting numerous training pairs is not an easy task considering HSIs. In \cite{zhang2021hyperspectral}, the authors collected real data for the training pair, but collecting these pairs, especially for remotely sensed HSI, requires extra effort and is an expensive procedure. A common approach used in HSI network training is to generate training pairs by adding synthetic noise to existing data \cite{zhang2019Hybrid,yuan2019Hyperspectral, maffei2020single, wei20203, wang2022sscan}. To obtain favorable outcomes, however, it is essential that the synthetic noise appearing in the training data possesses a distribution similar to the noise distribution in the \textcolor{black}{real-world} test data, which is hard to achieve in general. 

\textcolor{black}{Denoising HSIs affected by real noise poses a challenge due to the complex (non-normal) distribution of real noise statistics and its spatial and spectral variability. To tackle the complexities arising from diverse and intricate real noise, we present a dynamic and well-generalized denoising architecture. Dynamic neural networks, an emerging area in deep learning \cite{han2021dynamic, li2023spectral}, offer a distinct advantage over static models by adapting their structures or parameters to different inputs during inference. This adaptability enhances accuracy, representation power, and generality, all while maintaining computational efficiency.} 

\textcolor{black}{Our proposed dynamic deep neural network framework processes both spatial and spectral information of HSIs through feature-modulation layers, leveraging input neighboring spectral bands in an adaptive manner. At the heart of our approach lies the novel Spectral Self-Modulating Residual Block (SSMRB) used in the deep layers. The SSMRB intelligently regulates feature maps, preventing overfitting to the training set and elevating denoising performance. This mechanism yields two main benefits: (i) regulating the denoiser with spatial-spectral information, strengthening the network's learning ability, and (ii) establishing well-generalized denoising capabilities that effectively adapt to diverse noise patterns.} 

\textcolor{black}{Expanding on the novel SSMRB module, we also introduce the self-modulating CNN (SM-CNN) model for HSI denoising. Our experimental results demonstrate the effective generalization abilities of our proposed denoising method as compared to the existing studies. Surprisingly, even when trained with synthetic-noise data, our network achieves superior performance in denoising real-noisy HSIs. This outcome highlights the effectiveness and promise of our approach, empowered by the SSMRB, in handling real-noisy HSIs with high accuracy.} Our main contributions can be summarized as follows:

\begin{itemize}[leftmargin=*]
\item The core of our model is a new component, which we name as the spectral self-modulating residual block (SSMRB), that enables the network to adjust the features based on the adjacent spectral data \textcolor{black}{and prevents overfitting} to the training set, thus improving its capability to handle complex \textcolor{black}{real} noise.
\item The proposed SSMRB layer makes the proposed model a dynamic denoising network that adapts its predicted features according to the spatio-spectral attributes of each input HSI during the denoising process. This indicates that our network's weights can adapt themselves in real-time during the forward propagation, taking into account the spatio-spectral data.
\item By using the self-modulation framework, we have designed a new denoising network named SM-CNN, which efficiently removes noise from various HSIs that possess a varying number of spectral bands. It retains both the spectral information and local details of the HSI without requiring any additional parameter tuning, producing significantly more precise and clear outcomes.
 \item Our experimental analysis reveals that our approach outperforms the state-of-the-art HSI denoising methods both qualitatively and quantitatively on various noise cases including GN, SN, DN, IN, mixture \textcolor{black}{and real-world} noise in different spectra through a single SM-CNN model. Moreover, our proposed method allows for improving the classification performance on the real-world noisy HSI.
\end{itemize}

\section{Related Work}\label{sec:2}

\subsection{Classical Approaches to HSI Denoising}
In the last decade, the information that an HSI is well represented by a few pure spectral endmembers has been widely used. This property inspires researchers to exploit the low-rankness of HSI for denoising. In \cite{Zhang2014lrmr}, an HSI restoration technique based on the low-rank matrix recovery (LRMR) was developed. An extension of the LRMR in \cite{he2015hyperspectral} works well for removing mixed types of noise in HSI. With these methods, HSIs are rearranged as a 2D matrix and the low-rank feature of this matrix is examined. Other methods that exploit the low-rankness, such as LRTV \cite{he2015total}, LRTDTV \cite{wang2017hyperspectral} and 3D-GTV \cite{zhang20233d}, attempt to improve the performance of LRMR together with hyperspectral total variation (TV) \cite{yuan2012hyperspectral}. To further utilize low-rankness property, new models were proposed in \cite{zheng2020double, he2021tslrln, zeng2021hyperspectral, liu2022multi, zhang2022double, chen2022hyperspectral}. Majority of these methods can be considered as convex optimization problems and produce clean HSIs as global optima. However, the performance of these methods depends on the low-rankness prior that need to be investigated for each HSI, and especially TV-based ones cause over-smoothing. More comprehensive and in-depth information about model-based methods can be found in \cite{liu2023survey}, which provides a detailed overview of the different types of model-based methods and their strengths and weaknesses.

Another approach to HSI denoising is to use sparse representations \cite{peng2022low}. BM4D filtering \cite{maggioni2012bm4d}, an extension of the BM3D filter \cite{Dabov2007bm3d} to volumetric data, is an early example to these methods. BM4D groups 3D HSI patches and filters them in transfer domain. Non-local meets global (NGMeet) \cite{he2019non} offers a combined spatial non-local similarity and spectral low-rank approximation for HSI denoising. \textcolor{black}{FastHyDe \cite{zhuang2018fast} and FastHyMix \cite{zhuang2021fasthymix} make efficient use of both low-rankness and self-similarity properties of HSI for efficient denoising of GN and mixture noise, respectively.} Recently, SSSLRR \cite{zhao2022hyperspectral} proposes a denoising method based on both sparse representation and low-rankness. However, despite being suitable for Gaussian and Poisson noise, these sparse-coding based methods show poor performance in complex noise. 

\subsection{Deep Learning Approaches to HSI Denoising}
Deep learning can provide a prominent end-to-end learning strategy to solve the mentioned inadequacies of classical methods. To date, many DL methods for HSI denoising have been proposed. In the early days, methods originally developed for grayscale or RGB images were used for HSI denoising by adjusting the input and output filter sizes or treating them as a single band. DnCNN \cite{zhang2017beyond} suggested to use residual learning and batch normalization for fast convergence. HSI-DeNet \cite{chang2018hsi} introduced DL in HSI denoising based on a set of multi-channel 2D CNN filters for spatial and spectral structures. Moreover, MemNet \cite{tai2017memnet}, originally proposed for RGB denoising as a deep persistent memory network to solve long-term dependency issues, was used in \cite{wei20203} as a benchmark for noise removal in HSI with successful results. However, the approaches based on 2D filters cannot take full advantage of the substantial spectral information in HSIs.

Recently, HSI-specific networks have also been developed to take advantage of both the spectral and spatial properties of HSI \cite{wang2022sscan,wei20203,sidorov2019deep,miao2022hyperspectral,pan2022sqad, wang2022translution,xiong2022mac, lai2023mixed, hu2022hdnet}. In particular, SSCAN \cite{wang2022sscan} is an HSI-specific denoising network, that combines group convolutions and attention modules to effectively exploit spatial and spectral information in images. In \cite{wei20203}, a quasi-recurrent pooling function into the 3D (QRNN3D) U-net \cite{ronneberger2015u} was introduced to further capture the global correlation along the spectral spectrum. Such methods give state-of-the-art results, but since the images obtained from different sensors have variable spectral bands, the networks need to be reconstructed and retrained. Unfortunately, this is a time-consuming process. More recently, attention-based methods were proposed to capture non-local features \cite{pan2022sqad, wang2022translution,xiong2022mac, lai2023mixed}. 

Yuan \textit{et. al.} \cite{yuan2019Hyperspectral} proposed an HSI denoiser based on residual learning CNN (HSID-CNN), taking into account both spatial and spectral information and without the need to manually adjust hyperparameters for different HSIs. Similarly, a spatial-spectral gradient network (SSGN) \cite{zhang2019Hybrid} is proposed for the removal of mixed noise, using adjacent spectral difference in addition to spatial-spectral information. Maffei \textit{et. al.} developed a denoising method using a single CNN called HSI-SDeCNN~\cite{maffei2020single}. Since HSI-SDeCNN takes the noise variance as input, this model is effective for i.i.d.~GN. These methods yield the most successful results by training a single model. However, their success on test data depends on adding a synthetic noise to the training data similar to that in the test data. \textcolor{black}{As compared to these methods, our SM-CNN model can adapt itself during inference through the suggested SSMRB mechanism, which modulates the features for the network based on the provided spectral data. This also helps the suggested SM-CNN to partly alleviate the domain gap between the training and the test data distributions, as demonstrated in our experiments.}

\section{Proposed Denoising Method}\label{sec:3}

In this section, we describe our proposed single model self-modulating convolution neural network (SM-CNN) in detail.

\subsection{Hyperspectral Noise Model}
Considering HSI as a 3D tensor, assume that $ \mathbf{X} $ and $ \mathbf{Y} $ denote the clean data and the noisy observations, respectively, as given in Eqn.~\ref{eq:cleandata} and Eqn.~\ref{eq:noisydata}:

\begin{equation}\label{eq:cleandata}
\mathbf{X}=[\mathbf{X}_1,\mathbf{X}_2,...,\textcolor{black}{\mathbf{X}_i},...,\mathbf{X}_\mathit{B}]\in \mathbb{R}^{\mathit{M}\times\mathit{N}\times\mathit{B}}
\end{equation}
\begin{equation}\label{eq:noisydata}
\mathbf{Y}=[\mathbf{Y}_1,\mathbf{Y}_2,...,\textcolor{black}{\mathbf{Y}_i},...,\mathbf{Y}_\mathit{B}]\in\mathbb{R}^{\mathit{M}\times\mathit{N}\times\mathit{B}}
\end{equation}
where $\mathbf{X}_i$ and $\mathbf{Y}_i$ represent the clean and noisy grayscale images of $i$th band, $\mathit{M}\times\mathit{N}$ represents the spatial dimension and $\mathit{B}$ denotes the spectral dimension. An HSI degraded by additive sparse noise affecting some bands (\textcolor{black}{i.e.,} DN, SN, IN), $\mathbf{S}\in\mathbb{R}^{\mathit{M}\times\mathit{N}\times\mathit{B}}$, and dense noise affecting all bands (\textcolor{black}{i.e.,} GN), $\mathbf{N}\in\mathbb{R}^{\mathit{M}\times\mathit{N}\times\mathit{B}}$, can be linearly modeled as follows: $\mathbf{Y} = \mathbf{X} + \mathbf{S} + \mathbf{N} $. Hence, the HSI denoising process is the problem of estimating $\hat{\mathbf{X}}\in\mathbb{R}^{\mathit{M}\times\mathit{N}\times\mathit{B}}$,  which is an estimate of the $\mathbf{X}$, from noisy observation $\mathbf{Y}$.

\subsection{Method Description}\label{Sec:3-b}
\begin{figure*}[t]
\centering
\includegraphics[width=5.4in]{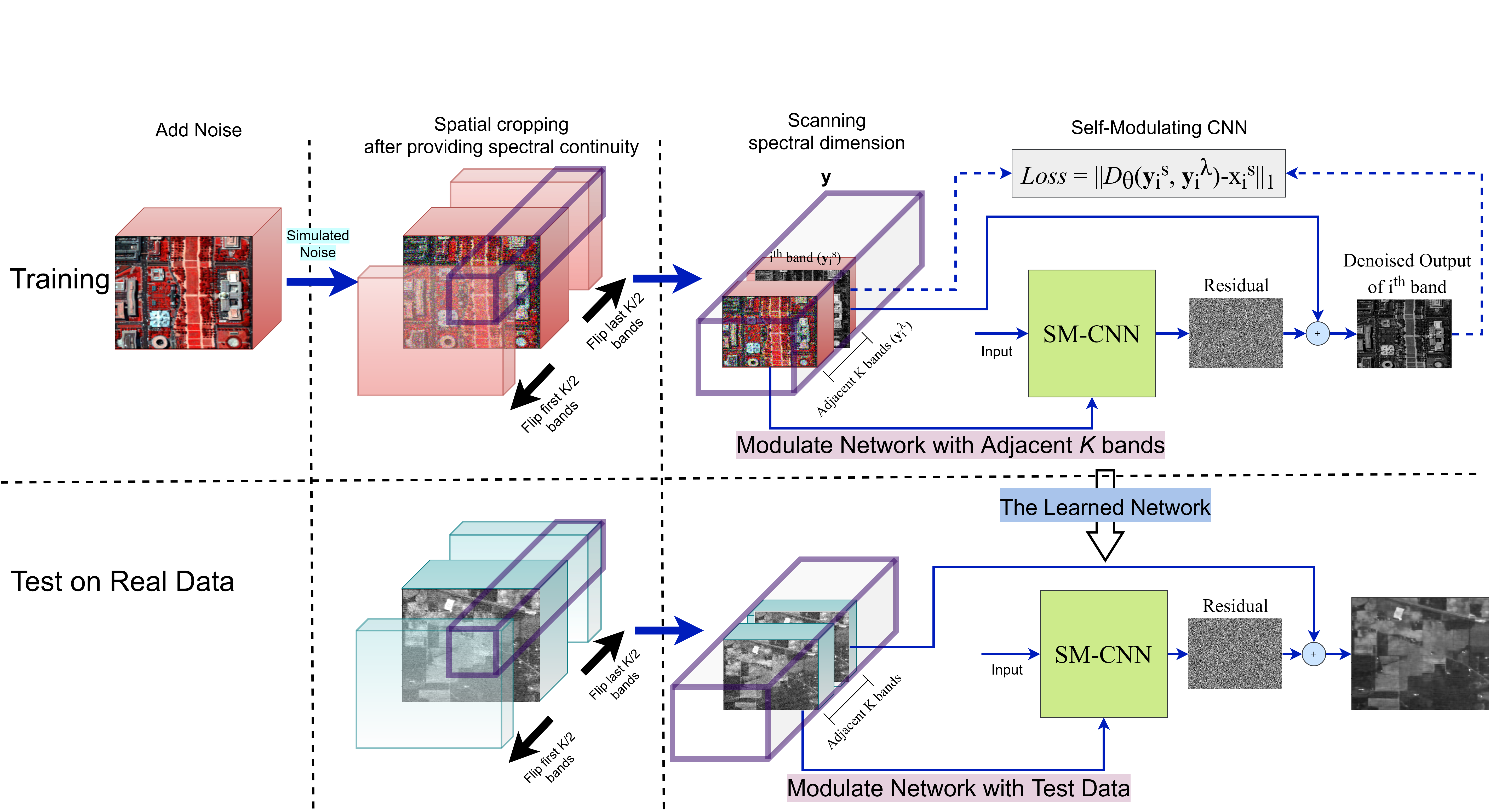}
\caption{System overview of the proposed SM-CNN model for HSI denoising. 
In training, we first obtain noisy input data by adding synthetic noise to clean HSI. Then, we perform spatial cropping after providing spectral continuity at the end points by flipping the $K/2$ bands. By continuously scanning the spectral dimension, we obtain adjacent $K$ spectral bands from the spatially cropped patches. We use spatially cropped neighbor bands as inputs when training the network to eliminate middle band noise. This process is repeated for all bands. The noisy $K$ adjacent bands as well as the input go through an SSMRB in deep layers to regulate the network, as given in Fig. \ref{fig_network}. Additionally, we use the residual learning strategy to ensure the stability and efficiency of the training by adding the middle band to the final output. In testing, we give input to the network after spatially cropping and spectrally scanning the data as in training. Here, the network is modulated with test correlated spectral bands through the SSMRB in the deep layers, \textcolor{black}{which lets the model adapt itself under complex noise settings not directly seen during training.}}

\label{fig_sm_cnn}
\end{figure*}
The system overview of the proposed method for HSI denoising is given in Fig. \ref{fig_sm_cnn}. Since HSIs are acquired by sensors that have different numbers of spectral bands and spatial diversity, a network typically needs to be retrained for different data. Hence, the strategy adopted to create a single model is to raster scan the spectral dimension and perform denoising one spatially cropped band at a time. With the SM-CNN method, a deep network structure is proposed to use both spatial and spectral information of HSI. When the proposed model performs the noise removal of a band, it uses the spatial information in the patch with size $h \times w$ pixels and the spectral correlation of $K$ adjacent bands including the target band. An important point is that we also use these $K$ neighbor bands to regulate our network  which is why we refer to it as self-modulating. In this way, we are able to significantly increase the noise reduction capacity and adaptability of the network for different types of noises (\textcolor{black}{i.e.,} GN, mixture noise etc.). Proposed training framework can be described as follows:

\begin{itemize}[leftmargin=*]
    \item[1)] Since SM-CNN comprehends a nonlinear peer-to-peer mapping between the noisy data and the clean data, we create training pairs by adding different synthetic noise to the original HSI data as shown in Fig. \ref{fig_sm_cnn}.
    \item[2)] We perform noise removal by scanning band-to-band spatial information and using $K$ adjacent spectral information along with spatial information as mentioned above. For this reason, we achieve spectral continuity at the end points by flipping the $K/2$ bands.
    \item[3)] Then, we obtain patches $\mathbf{y}$ of size $h\times w\times (B+K)$ by spatially cropping the enlarged noisy data of size $M\times N\times (B+K)$. 
    \item[4)] Lastly, we get adjacent $K$ spectral bands from spatially cropping patches by scanning the spectral dimension continuously. For $i\in [1, B]$, we use one of the middle bands $\mathbf{y}_{i}^{s}$ and its neighboring bands $\mathbf{y}_{i}^{\lambda}$ as inputs to the network to eliminate noise of the middle band $\mathbf{y}_{i}^{s}$. The network is trained with this middle band $\mathbf{y}_{i}^{s}$ and its corresponding clean target $\mathbf{x}_{i}^{s}$. At the same time, these neighboring bands are used for spectral self-modulation of the network.
    \item[5)] For each spatially cropped patch $\mathbf{y}$, the input $\mathbf{y}_{i}^{s}$, its corresponding spectral bands $\mathbf{y}_{i}^{\lambda}$, and clean target $\mathbf{x}_{i}^{s}$ are selected and this process is repeated for all bands.
\end{itemize}

After the training is completed, the test process is performed as shown in Fig. \ref{fig_sm_cnn}. Firstly, just like in training, we provide spectral continuity at the end points by flipping the $K/2$ bands of the test data. The enlarged test data is then cropped in the same way as in training. As shown in Fig. \ref{fig_sm_cnn}, the network is modulated by the neighboring bands of test data. This allows the network to give much better denoising performance.

\subsection{Network Architecture}
\begin{figure*}[t]
\centering
\includegraphics[width=5.4in]{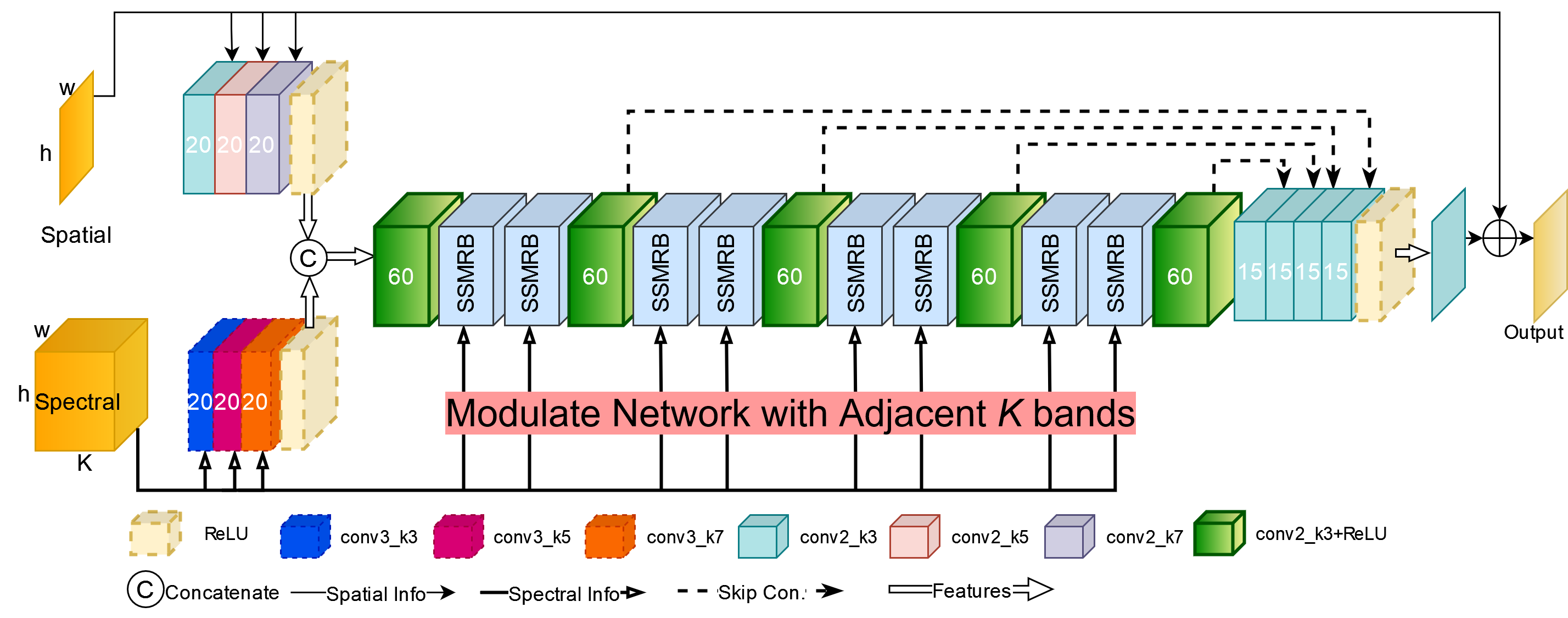}
\caption{Structure of the proposed SM-CNN. The SSMRB used in the deep layers is given in Fig. \ref{fig_ain} and detailed in Sec. \ref{sec:3-1}.}
\label{fig_network}
\end{figure*}
The overall architecture of the SM-CNN model is shown in Fig. \ref{fig_network}. The numbers written on the convolution layers in the figure indicate the number of channels. Our network has two input data. The input spatial data of size $h\times w$ represents the current noisy band to be denoised, and the input spectral data of size $h\times w\times K$ represents the spatial-spectral information of adjacent bands. As we mentioned above, based on the learning strategy of the proposed network, different HSI data can be denoised with a single model regardless of the band number, because a patch-based band-to-band denoising process is performed. One of the most important points here is that we use the provided correlated spectral data through the SSMRB in the deep layers to modulate the denoiser. The benefits are two folds. First, it regulates the denoiser with spatial-spectral information, thus increasing the learning ability of the network. Second, it enables the denoiser to generalize better \textcolor{black}{through the feature modulation process, which lets the model to adapt itself for the novel noise settings.} These gains are achieved by normalizing the features at deep layers according to the spectral information. In deep layers, normalization parameters of the features are obtained from the pixel-by-pixel noise level. Then, normalized feature are scaled and shifted according to the input adjacent spectral data, making them context dependent.

Our network's core structure is inspired by the HSID-CNN \cite{yuan2019Hyperspectral} in which the given hyperspectral image is first processed with both 2D and 3D CNN layers to encode spatial and spectral information. \textcolor{black}{However, the introduction of the SSRMB block in the architecture brings a significant difference in the behavior of our model, particularly in handling unseen noise distributions. The SSRMB block empowers the network to adapt itself to data with characteristics completely different from those seen during training. This adaptability is crucial in real-world scenarios where noise distributions can vary widely.} 

In particular, as shown in Fig.~\ref{fig_network}, a single 2D spatial spectral band input is processed with 2D CNN layers with different kernel sizes (conv2\_k3, conv2\_k5, conv2\_k7 where k stands for kernel size).  Then, in parallel, a 3D spatio-spectral hypercube input is processed with 3D CNN convolutional layers with different kernel sizes (conv3\_k3, conv3\_k5, conv3\_k7). Finally, these 2D and 3D convolutional features are concatenated. These 2D and 3D CNN structured at the initial stage are considered in order to better use and investigate the characteristics of a single band and to make high use of adjacent correlated spectral bands. After applying rectified linear unit (ReLU) to conjoined outputs of 2D CNN and 3D CNN, these are concatenated and forwarded to sequential deep layers consisting of 2D CNN followed by ReLU blocks (conv2\_3k+ReLU) and the proposed SSMRB layers. \textcolor{black}{To strike the right balance between model performance and computational efficiency, the number of consecutive SSMRBs is determined through careful experimental selection.} As will be detailed in Sec.~\ref{sec:3-1}, the SSMRB module make the network aware of the given spectral signal within every step of the denoising process, and highly improves the performance of the core spectral-spatial denoising network as we demonstrate in our experiments. We use skip connections from sequential deep layers to output layer in order to ensure training stability \cite{zhang2017beyond}. These four skip connection are passed through the 2D CNN with 15 channels (conv2\_k3), and then all outputs are concatenated. Lastly, final layer of the proposed denoiser is a single channel 2D CNN to obtain the estimation of the clean data from the noisy spatial channel.

\subsection{Spectral Self-Modulating Residual Block}\label{sec:3-1}
\begin{figure*}[htb!]
\centering
\includegraphics[width=3.75in]{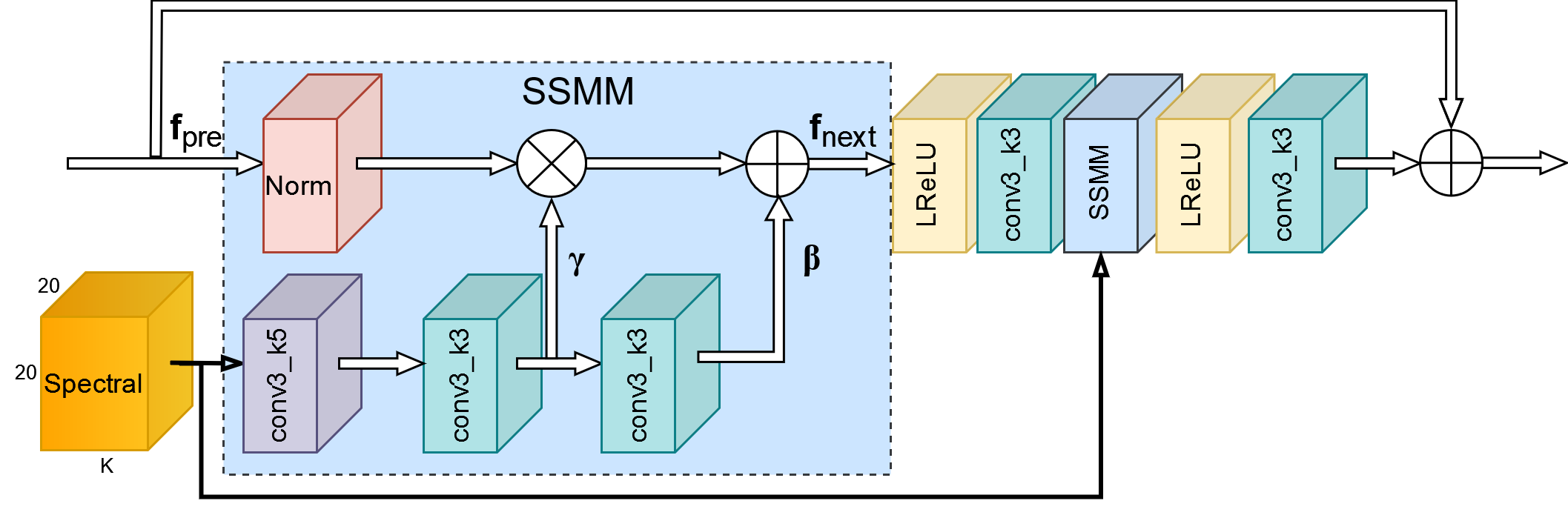}
\caption{Spectral self-modulation residual block (SSMRB) used in our SM-CNN denoiser network. This block modulates features in deep layers by using spectral neighbor bands for adapting the network to noise.}
\label{fig_ain}
\end{figure*}

We modulate our network with SSMRB by using the spectral signal itself, and thus we call our network as self-modulating CNN. The structure of the SSMRB is displayed in Fig. \ref{fig_ain}. This block consists of two spectral self-modulation modules (SSMM) and two 2D CNNs connected consecutively. The SSMM transmutes the previous feature map $\mathbf{f}_\mathrm{pre}\in\mathbb{R}^{\mathit{h}\times\mathit{w}\times\mathit{C}}$ by taking input adjacent spectral information where $h\times w$ denotes the spatial size of the feature map, and C is the number of channels. To produce the affine transformation of the feature maps, the SSMM first normalizes the feature map channel-wise and then generates scale ($\mathbf{\gamma}$) and shift ($\mathbf{\beta}$) for each pixel by using the adjacent spectral bands, giving the transformed features:
\begin{equation}\label{eq:feature_norm}
    \mathbf{f}_\mathrm{next}^c = \mathbf{\gamma}_c(\mathbf{y}^{\lambda})\frac{\mathbf{f}_\mathrm{pre}^c-\mu_c}{\sigma_c} + \mathbf{\beta}_c(\mathbf{y}^{\lambda})
\end{equation}
where $\mathbf{\gamma}_c(\mathbf{y}^{\lambda})$ and $\mathbf{\beta}_c(\mathbf{y}^{\lambda})$ are the learned self-modulation parameters obtained pixel-wisely from the input spectral bands $(\mathbf{y}^{\lambda})$ for each $c \in[1,C]$ with $C=60$. $\mu_c$ and $\sigma_c$ refer to the mean and standard deviation of feature map $\mathbf{f}_\mathrm{pre}^c$ for channel $c$, respectively; and can be formulated as:
\begin{equation}
    \mu_c = \frac{1}{hw}\sum_{l}^{h}\sum_{k}^{w} \mathbf{f}_\mathrm{pre}^c(l,k)
\end{equation}
\begin{equation}
    \sigma_c = \sqrt{\frac{1}{hw}\sum_{l}^{h}\sum_{k}^{w} (\mathbf{f}_\mathrm{pre}^c(l,k)-\mu_c)^2 + \delta}
\end{equation}
with $\delta$ denoting a stability parameter to prevent Eqn. \ref{eq:feature_norm} from dividing by zero. We set $\delta=10^{-5}$ for our SM-CNN.

\textcolor{black}{From a theoretical point of view, there are some parallels that can be drawn between the proposed SSMRB block and the attention blocks, which have been a core building block in contemporary neural networks. However, it is essential to disambiguate the differences between the two. The distinction lies in the distinct intuitions underlying the operation of attention and the feature modulation scheme in our SSMRB block. In particular, attention mechanisms assume that specific spatial locations or feature channels contain the most useful information and select these locations or channels for further processing. On the other hand, the SSMRB block performs a spatially-varying affine transformation on the extracted feature maps, conditioned on the characteristics of the adjacent spectral data. It considers the correlations and relationships between neighboring spectral bands and modulates the features accordingly. In this sense, the SSMRB block can be seen as a special kind of normalization layer, where the normalization factors representing mean and variance are estimated based on the adjacent spectral data during inference time. This adaptability allows the network to adjust itself to the specific spectral characteristics of the input data, making it more robust to different noise distributions.}

\subsection{Loss Function}
In training our proposed model, we employ the residual learning strategy to ensure the stability and efficiency of the training procedure \cite{yuan2019Hyperspectral, zhang2017beyond}. Given a training set 
$\{(\mathbf{y}_{i}^{s}, \mathbf{y}_{i}^{\lambda}), \mathbf{x}_{i}^{s}\}_{i=1}^{N}$ 
where $N$ is the number of training patches, $\mathbf{x}_{i}^{s}$ is a single-band clean patch of noisy low-quality patch $\mathbf{y}_{i}^{s}$, and $\mathbf{y}_{i}^{\lambda}$ is the noisy $K$ adjacent spectral bands of $\mathbf{y}_{i}^{s}$. The loss function of the proposed SM-CNN denoiser ($D_\theta$) with the parameter set $\theta$ is:

\begin{equation}\label{eq:loss_function}
    \mathcal{L}(\theta)= \frac{1}{2N}\sum_{i=1}^{N}\|D_\theta(\mathbf{y}_{i}^{s}, \mathbf{y}_{i}^{\lambda})-\mathbf{x}_{i}^{s}\|_{1}
\end{equation}

\section{Experimental Results}\label{sec:4}
We have evaluated the effectiveness of the proposed SM-CNN method using both synthetic noisy and real noisy HSIs. First, effectiveness of the method has been verified using simulated data by adding synthetic noise. Then the proposed method has been applied to real noisy images. The proposed method has been compared with the classical approaches of tensor TDL \cite{peng2014TDL}, LRTF-DFR \cite{zheng2020double}, FastHyMix \cite{zhuang2021fasthymix}, BM4D \cite{maggioni2012bm4d}, LRMR \cite{Zhang2014lrmr}, LRTV \cite{he2015total}, and LRTDTV \cite{wang2017hyperspectral}, for which codes are publicly available. In the field of DL, we have compared the proposed method with HSID-CNN \cite{yuan2019Hyperspectral}, MemNet \cite{tai2017memnet}, QRNN3D \cite{wei20203}, \textcolor{black}{HDNET} \cite{hu2022hdnet} and \textcolor{black}{MAN} \cite{lai2023mixed}. For a fair comparison, we have trained a version of MemNet, which has 6 memory blocks and 6 residual blocks, like the proposed method. For this, we have set the input layer filter to $K$, the output layer to 1 and perform the training as suggested in Sec. \ref{Sec:3-b}. \textcolor{black}{Additionally, we have performed training for HDNET and MAN using the WDC dataset.} On the other hand, since Wie \textit{et. al.} \cite{wei20203} trained the QRNN3D network using synthetic noise similar to our cases, we have obtained results using pre-trained networks as a single model for a fair comparison. Here, it should be noted that QRNN3D can achieve better results if both the train and the test set are of similar content, as detailed in \cite{wei20203}.

In addition to visual comparison, three commonly used metrics have been adopted to evaluate the performance of the proposed approach on simulated data: mean peak signal-to-noise-ratio (MPSNR), mean structural similarity index (MSSIM), and spectral angle mapper (SAM). MPSNR and MSSIM show the spatial accuracy, which are calculated on each 2D spatial image and averaged over all spectral bands. SAM that shows the spectral fidelity is calculated on each 1D spectral vector and averaged over all spatial points. The higher the MPSNR and MSSIM scores and the lower the SAM score mean the better denoising results. To further evaluate the effectiveness of the proposed model, two real-world noisy HSI datasets have been used in our real-data experiments. Moreover, the performance of the methods has been also evaluated by performing a classification task on a real noisy HSI. First, we apply the denoising model to the data, then employ a support vector machine (SVM) \cite{archibald2007feature} as the classifier before and after denoising. Finally, overall accuracy (OA) and kappa coefficient are given as evaluation indexes. 
 
\subsection{Datasets}\label{sec:4-a}
Four HSI datasets are considered to evaluate the effectiveness of the proposed method: one is used to train the network and conduct experiments by introducing different complex simulated noise, while the other three are used in complex simulated noise and real noise experiments, to evaluate the performance of the proposed model.

\noindent \textbf{Training Dataset.} To train the proposed model and other DL methods, we use a part of the Washington DC Mall (WDC\footnote{\url{https://engineering.purdue.edu/~biehl/MultiSpec/hyperspectral.html}}) data acquired by the Hyperspectral Digital Image Acquisition Experiment (HYDICE) airborne sensor \cite{wang2014wdc}. The sensor system records 210 spectral bands in the 0.4 to 2.4 $\mu$m region of the visible and infrared spectrum. The bands in the 0.9 and 1.4 $\mu$m regions in which there is atmospheric interference have been removed from the dataset. Hence, spatial resolution of the WDC data is 1208$\times$307$\times$191 pixels. We divide this data into two parts. One part with 200$\times$200$\times$191 pixels is used for testing, and the remaining part is used for training and validation.

\noindent \textbf{Testing Datasets.} Four datasets have been used in simulated and real data experiments to evaluate the effectiveness of the proposed method as follows:

\begin{itemize}[leftmargin=*]
    \item\textit{Washington DC Mall:} As mentioned above, an area of 200$\times$200$\times$191 pixels is used for simulated data experiments by adding synthetic noise to the original image.
    \item\textit{Pavia University (PU\footnote{\url{http://lesun.weebly.com/hyperspectral-data-set.html}}):} This data was acquired by the Reflective Optics Spectrographic Imaging System (ROSIS) over Pavia, Italy. The  scene of size 200$\times$200$\times$103 with a spectral range from 0.43 to 0.86 $\mu$m is used in the simulated data experiment by introducing mixture noise after removing the 12 water absorption bands.
    \item \textit{Indian Pines (IP\footnote{\url{https://purr.purdue.edu/publications/1947/about?v=1}}):} IP data was collected by the Airborne Visible Infra-Red Imaging Spectrometer (AVIRIS) over the Indian Pines test site, and consists of 145$\times$145 pixels and 224 spectral bands in the wavelength range 0.4–2.5 $\mu$m \cite{baumgardner2015220}. After removing the bands severely damaged by the atmosphere and water (150-163), a total of 206 bands are used in the experiments.
    \item \textcolor{black}{\textit{HSIDwRD\footnote{\url{https://github.com/ColinTaoZhang/HSIDwRD}}:} This data was acquired by a SOC710-VP hyperspectral camera with 696$\times$520 pixels in spatial resolution and 34 spectral bands from 0.4 $\mu$m to 0.7 $\mu$m \cite{zhang2021hyperspectral}. The dataset contains 59 clean images captured using long-exposure settings and their corresponding noisy images captured using short-exposure settings.}
\end{itemize}
 
WDC and PU are assumed to be noise-free, and used in testing by adding synthetic noise. IP and HSIDwRD, on the other hand, are used directly in testing since many of their bands are inherently noisy.

\subsection{Experimental Settings}
In this section, we describe the implementation details. First, HSI data is scaled to [0, 1] before adding noise during training of simulated-data experiments. Then, noise components are generated for the different cases given in the following subsection and added to the original data. As described in \ref{Sec:3-b}, when the SM-CNN performs the denoising of a band, it uses the spatial information in the patch with size $h\times w$ and the spectral correlation of $K$ adjacent bands including the target band. We set the patch size to $20\times20$ with the stride equal to $10$, and rotate the patches $0^{\circ}$, $90^{\circ}$, $180^{\circ}$ and $270^{\circ}$ for data augmentation during training. \textcolor{black}{Based on experimental findings, the number of adjacent spectral bands is fixed at $K = 24$.} 

\noindent \textbf{Simulated Noise Settings.} HSIs are usually degenerated by different noise types including  GN, SN, DN, IN and their combinations. Therefore, following the experimental setting used in~\cite{wei20203}, we define five types of complex noise added to simulated data for training and testing purposes as follows:
\begin{itemize}[leftmargin=*]
	\item \textbf{Case 1}: Non-i.i.d. GN. All bands are corrupted by zero-mean GN with a random intensity ranged from 10 to 70.
	\item \textbf{Case 2}: GN \& SN. All bands are corrupted by non-i.i.d. GN as Case 1. In addition, one third of bands (63 bands for WDC datasets) are randomly selected to add SN of $ 5\% $ to $ 15\% $ of columns.
	\item \textbf{Case 3}: GN \& DN. The noise generation process is similar to Case 2, except that the SN is replaced by DN.
	\item \textbf{Case 4}: GN \& IN. Each band is polluted by GN as Case 1. Besides, one third of bands are randomly chosen to add IN with intensity ranged from $10\%$ to $70\%$.
	\item \textbf{Case 5}: Mixture noise. Each band is randomly corrupted
    by GN as Case 1 and at least one type of noise specified in Cases 2-4.
\end{itemize}
\noindent \textbf{Training Details.} SM-CNN is implemented with PyTorch and trained on NVIDIA TESLA V100 GPUs. It is trained by minimizing the mean absolute error loss (MAELoss) between the noisy HSI patches and the corresponding clean patches. The network parameters are initialized with the XavierNormal initializer and updated by the Adam optimizer. The learning rate is set to $0.0001$ and the mini-batch to $128$. SM-CNN's training period took $100$ epochs and the best performance on the validation data was recorded. Lastly, the training time for each model was approximately 4-5 hours.

\subsection{Simulated-Data Experiments}

In this section, we present both quantitative and visual results of our proposed network on simulated test data as mentioned above. The proposed model is discussed by comparing it with those in the literature. 

\noindent \textbf{WDC dataset.} We first train our network for 5 complex noise cases and test it via the WDC dataset. Table~\ref{table:wdc} lists the quantitative comparisons of competing methods obtained with WDC data for cases 1-5. Each column in the table shows the metric results obtained with different noise cases. For MPSNR and MSSIM, the higher the better; for SAM, the lower the better. While the noisy HSI row shows the metric values of the synthetic noisy data for different cases, the other rows show the metric values obtained with the results of the classical methods, deep methods in the literature and the proposed method, respectively. The best performance for each quality index in the table is marked in bold, and the second best performance for each quality index is marked with underline.

\renewcommand*{\arraystretch}{1.45}
\begin{table*}[hbt!]
  \centering
  \caption{Quantitative evaluation of different denoising methods with five complex noise cases on the WDC dataset.}
    \scalebox{0.7}{\begin{tabular}{l|c@{$\;$}c@{$\;$}c|c@{$\;$}c@{$\;$}c|c@{$\;$}c@{$\;$}c|c@{$\;$}c@{$\;$}c|c@{$\;$}c@{$\;$}c}
\toprule
\multirow{2}{*}{Method} & \multicolumn{3}{c|}{Case 1: GN} & \multicolumn{3}{c|}{Case 2: GN \& SN} & \multicolumn{3}{c|}{Case 3: GN \& DN} & \multicolumn{3}{c|}{Case 4: GN \& IN} & \multicolumn{3}{c}{Case 5: Mixture Noise} \\
 & {MPSNR$\uparrow$} & {MSSIM$\uparrow$} & SAM$\downarrow$ & {MPSNR$\uparrow$} & {MSSIM$\uparrow$} & SAM$\downarrow$   & {MPSNR$\uparrow$} & {MSSIM$\uparrow$} & SAM$\downarrow$   & {MPSNR$\uparrow$} & {MSSIM$\uparrow$} & SAM$\downarrow$   & {MPSNR$\uparrow$} & {MSSIM$\uparrow$} & SAM$\downarrow$ \\
\midrule
Noisy HSI & 18.508 & 0.690 & 0.278 & 18.982 & 0.711 & 0.264 & 17.338 & 0.653 & 0.328 & 15.269 & 0.531 & 0.420 & 13.402 & 0.500 & 0.464 \\
BM4D \cite{maggioni2012bm4d}  & 26.904 & 0.943 & 0.093 & 27.057 & 0.947 & 0.091 & 23.303 & 0.895 & 0.158 & 20.272 & 0.729 & 0.251 & 17.841 & 0.701 & 0.297 \\
LRTV \cite{he2015total} & 25.464 & 0.906 & 0.111 & 25.839 & 0.914 & 0.106 & 24.895 & 0.895 & 0.118 & 23.253 & 0.850 & 0.153 & 22.842 & 0.855 & 0.150 \\
LRMR \cite{Zhang2014lrmr} & 28.501 & 0.964 & 0.079 & 28.670 & 0.965 & 0.077 & 25.926 & 0.944 & 0.104 & 23.183 & 0.865 & 0.170 & 21.841 & 0.864 & 0.173 \\
LRTDTV \cite{wang2017hyperspectral} & 27.999 & 0.956 & 0.081 & 28.376 & 0.960 & 0.077 & 27.602 & 0.952 & 0.084 & 26.393 & 0.933 & 0.100 & 26.041 & 0.931 & 0.101 \\
LRTF-DFR\cite{zheng2020double} & 31.603  & 0.981 & 0.054 & 31.987 & \underline{0.983} & \textbf{0.051} & 30.621 & 0.978 &0.058  &28.988 &0.964  & 0.081 & \underline{28.559} & \underline{0.967} & \underline{0.077} \\
FastHyMix\cite{zhuang2021fasthymix} & \underline{32.303} & \textbf{0.986}  & \underline{0.052}  & \textbf{32.155} & \textbf{0.986} & \underline{0.052}  & 30.203 & \underline{0.981} & 0.059 &27.435 & 0.911 & 0.129 & 24.618 & 0.905 & 0.133\\
QRNN3D \cite{wei20203} & 27.352 & 0.963 & 0.084 & 27.512 & 0.965 & 0.082 & 27.336 & 0.964 & 0.084 & 26.943 & 0.960 & 0.088 & 26.197 & 0.952 & 0.096 \\
HSID-CNN \cite{yuan2019Hyperspectral} & 29.355 & 0.968 & 0.071 & 29.541 & 0.970 & 0.069 & 28.872 & 0.966 & 0.075 & 26.559 & 0.943 & 0.097 & 26.156 & 0.940 & 0.101 \\
MemNet \cite{tai2017memnet} & 28.126 & 0.964 & 0.095 & 28.398 & 0.966 & 0.079 & 29.913 & 0.971 & 0.069 & 29.702 & 0.969 & 0.081 & 27.082 & 0.960 & 0.093 \\
\textcolor{black}{HDNET} \cite{hu2022hdnet} & 29.897 & 0.972 & 0.065 & 30.079 & 0.974 & 0.064 & 29.158 & 0.969 & 0.070 & 26.982 & 0.951 & 0.090 & 26.222 & 0.945 & 0.095 \\
\textcolor{black}{MAN} \cite{lai2023mixed}  & 31.971 & 0.981 & 0.052 & \underline{32.060} & \underline{0.983} & \textbf{0.051} & \underline{31.664} & \underline{0.981} & \underline{0.054} & \underline{29.973} & \underline{0.971} & \underline{0.066} & 28.205 & 0.961 & 0.079\\
SM-CNN (Ours) & \textbf{32.529} & \underline{0.984} & \textbf{0.048} & 31.477 & 0.981 & 0.054 & \textbf{32.281} & \textbf{0.983} & \textbf{0.050} & \textbf{30.063} & \textbf{0.973} & \textbf{0.064} & \textbf{29.832} & \textbf{0.973} & \textbf{0.066} \\
    \bottomrule
    \end{tabular}
    }%
  \label{table:wdc}%
\end{table*}%

 \begin{figure*}[htb!]
\centering
\includegraphics[width=5.5in]{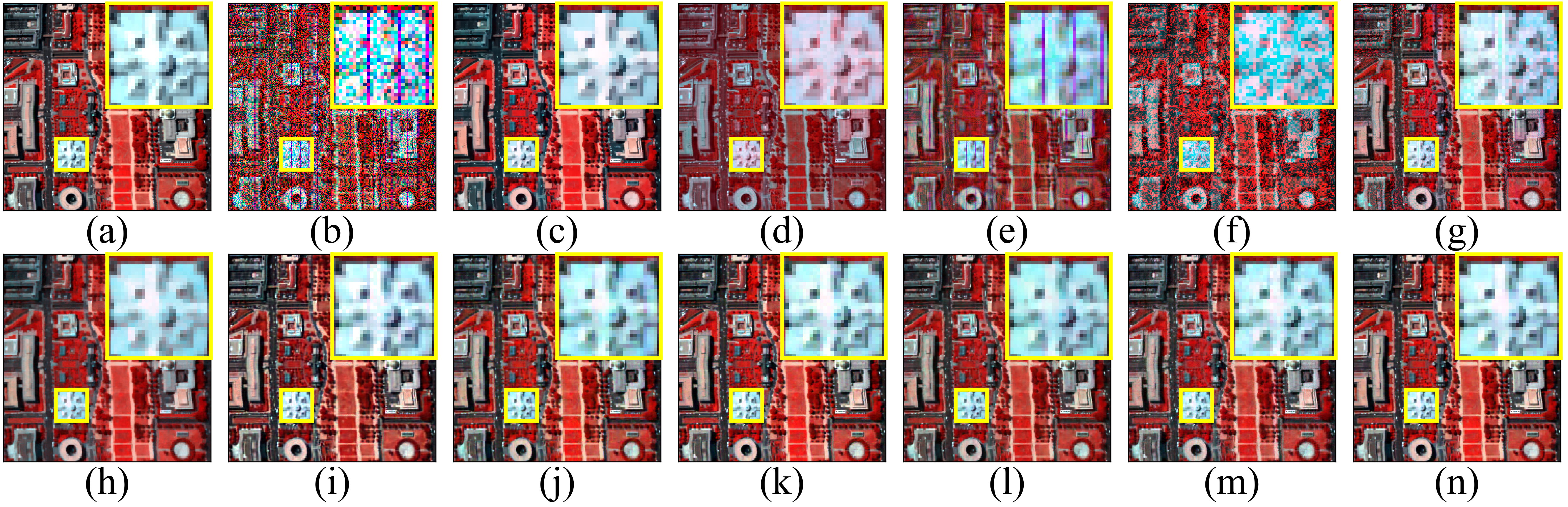}
\caption{Results for WDC with mixture noise in Case 5. (a) False-color original image with bands (57, 27, 17), (b) Noisy image, (c) LRTF-DFR, (d) FastHyMix, (e) BM4D, (f) LRTV, (g) LRMR, (h) LRTDTV, (i) QRNN3D, (j) HSID-CNN, (k) MemNet, (l) HDNET, (m) MAN, (n) SM-CNN (Ours).}
\label{fig_wdc_case5}

\end{figure*}

Additionally, to make visual comparisons for Case 5, we have selected bands 57, 27 and 17 to generate false-color images of WDC data obtained by all methods and they are shown in Fig. \ref{fig_wdc_case5}. Specifically, Fig. \ref{fig_wdc_case5}(a) shows the original image and Fig. \ref{fig_wdc_case5}(b) demonstrates noisy image before denoising, while Fig. \ref{fig_wdc_case5}(c)–(n) shows the images obtained after applying different denoising methods. Moreover, to examine figures in more detail, we have zoomed in on a region and showed this region in upper right part of the figures.

\textcolor{black}{As can be seen in Table \ref{table:wdc}, the proposed SM-CNN achieves the highest MPSNR and MSSIM values and the lowest SAM values at four out of five complex cases. Especially in the mixture noise case, the proposed method has increased the level of success compared to both classical and deep methods. Looking at the traditional methods, BM4D performed quite well for Case 1 and Case 2 on the basis of metrics because it is an effective model for GN removal. However, BM4D in particular causes the image to become smooth and reduces edge details. As the noise complexity increases, the performance of BM4D drops dramatically. FastHyMix demonstrated impressive performance in the first three cases, particularly in case 2, but its effectiveness significantly declines in mixture noise. This method introduces artifacts and does not handle complex noise case as shown in Fig. \ref{fig_wdc_case5}. On the other hand, LRTV, LRTDTV and LRTF-DFR produced better results in complex case among others conventional methods, since they are more suitable for these situations. But, TV norm-based LRTV and LRTDTV methods show the smoothing effect. LRMR, in contrast, suffers from artifacts in the mixture case which are evident in Fig. \ref{fig_wdc_case5}. LRTF-DFR produced fairly good results in most situations. However, these methods require the careful adjustment of several hyper-parameters. While generating these results, we tried our best to produce the best results by adjusting the parameters.}

As can be seen in Fig. \ref{fig_wdc_case5}, the results of the DL methods are visually close, but the proposed method appears to be good at recovering the details. Also, the quantitative analysis in Table \ref{table:wdc} reveals that our method outperforms in all metrics for all cases except case 2. In this table, higher performance on the MSSIM metric for each noise case indicates that the proposed model has a stronger and more robust ability to preserve structure properties and recover edge and detail information. Further, the superior performance on the SAM metric proves that SM-CNN can better maintain spectral accuracy than other methods.

\begin{figure*}[htb!]
\centering
\includegraphics[width=5.4in]{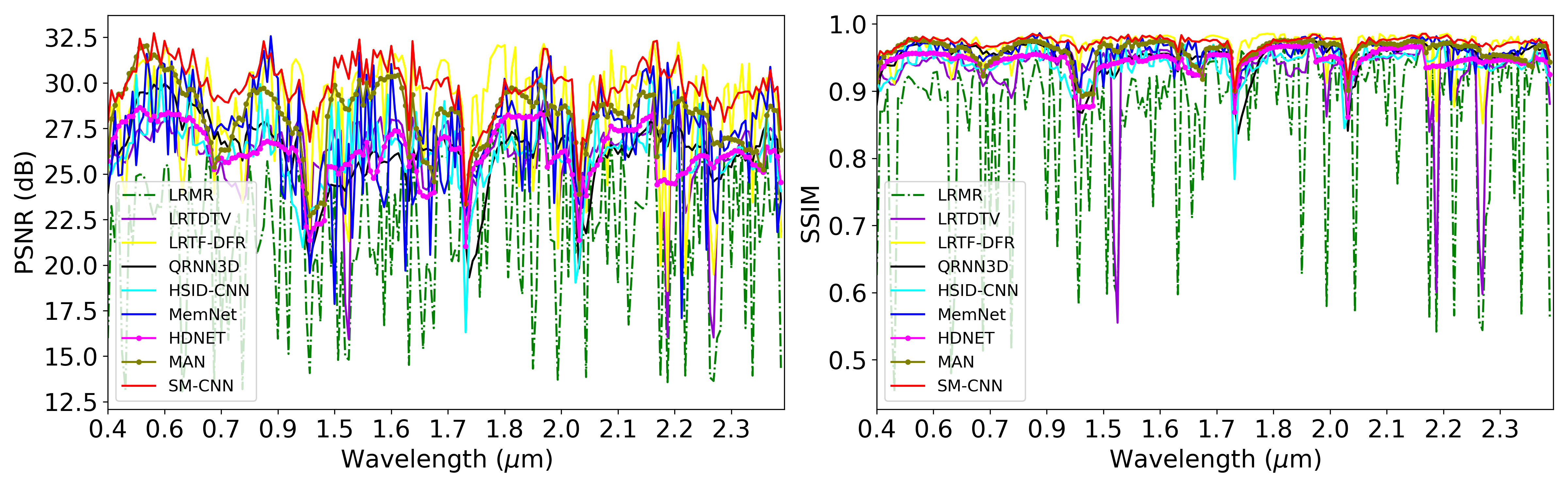}
\caption{PSNR and SSIM values across the spectrum corresponding to the denoising results of the proposed and the competing methods for Case 5.}
\label{fig_psnr_ssim}
\end{figure*}

\textcolor{black}{Fig. \ref{fig_psnr_ssim} shows the denoising results with the PSNR and SSIM values for simulated mixture noise case across the spectral spectrum. Here, we present the outcomes of LRTDTV and LRTF-DFR which give the best results among traditional methods, and LRMR that inspired many of the classical methods. Additionally, we show the results of the deep models. The values given in Table \ref{table:wdc} are obtained by averaging the PSNR and SSIM values in Fig. \ref{fig_psnr_ssim} along the spectrum. In general, our method outperforms the others for almost all bands in terms of both PSNR and SSIM metrics. While the performance of LRMR fluctuates between bands, LRTDTV and LRTF-DFR produce more stable results across the spectral spectrum. Nonetheless, in certain wavelengths (\textcolor{black}{i.e.,} around 2.2 $\mu$m and 2.3 $\mu$m), the efficacy of LRTDTV and LRTF-DFR seems to decrease, as evidenced in Fig. \ref{fig_psnr_ssim} by the decline in both PSNR and SSIM values.} MemNet and HSID-CNN approaches have a performance that fluctuates between the bands, but not as much as LRMR. Our method outperforms MemNet in most of the bands. \textcolor{black}{HDNET and MAN methods show a decrease in their performance compared to our method at points where spectral continuity is disrupted.} QRNN3D method obtains more stable results across the spectral spectrum, but it needs to be retrained for better results rather than considering it as a single model, as noted in \cite{wei20203}.

\noindent \textbf{PU dataset.} The noises similar to the mixture noise given in Case 5 have been added to this data. All conditions except the randomly selected number of bands to add different noises have been chosen as in the above cases. Since the number of bands is less than those of WDC, one third of the randomly selected bands corresponds to 34 bands. As a result, each band is randomly corrupted by at least one type of noise. While performing this test, the network trained on WDC data has been used. Note that the pre-trained network is used even though the number of bands is different. This feature demonstrates the reusability power of single models.

\renewcommand*{\arraystretch}{1.00}
\begin{table}[hbt!]
  \centering
  \caption{Quantitative evaluation of different denoising methods on the PU dataset across 10 noisy runs.}
    \scalebox{0.55}{\begin{tabular}{p{0.22\textwidth}>{\centering}p{0.18\textwidth}>{\centering}p{0.18\textwidth}>{\centering\arraybackslash}p{0.18\textwidth}}
    \toprule
    Method & MPSNR$\uparrow$ & MSSIM$\uparrow$ & SAM$\downarrow$ \\
    \midrule
    Noisy HSI & 14.671$\pm$0.317 & 0.281$\pm$0.015 & 0.649$\pm$0.012\\
    BM4D \cite{maggioni2012bm4d}  & 23.853$\pm$0.212 & 0.713$\pm$0.010 & 0.273$\pm$0.008 \\
    LRTV \cite{he2015total}  & 27.786$\pm$1.005 & 0.838$\pm$0.015 & 0.204$\pm$0.023 \\
    LRMR \cite{Zhang2014lrmr}  & 26.844$\pm$0.280 & 0.809$\pm$0.011 & 0.214$\pm$0.007 \\
    LRTDTV \cite{wang2017hyperspectral} & 31.038$\pm$0.191 & 0.901$\pm$0.006 & 0.142$\pm$0.005 \\
    LRTF-DFR\cite{zheng2020double} & 29.696$\pm$0.255  & 0.902$\pm$0.007 & 0.138$\pm$0.005  \\
    FastHyMix\cite{zhuang2021fasthymix} & 28.370$\pm$0.290  & 0.884$\pm$0.007 & 0.162$\pm$0.007 \\
    QRNN3D \cite{wei20203} & \underline{31.274$\pm$0.134} & \textbf{0.953$\pm$0.002} & \textbf{0.107$\pm$0.002} \\
    HSID-CNN \cite{yuan2019Hyperspectral} & 27.185$\pm$0.105 & 0.837$\pm$0.003 & 0.189$\pm$0.002 \\
    MemNet \cite{tai2017memnet} & 29.642$\pm$0.118 & 0.910$\pm$0.004 & 0.143$\pm$0.001 \\
    \textcolor{black}{HDNET} \cite{hu2022hdnet} & 29.931$\pm$0.186 & 0.911$\pm$0.004 & 0.142$\pm$0.003 \\
    \textcolor{black}{MAN} \cite{lai2023mixed} & 30.283$\pm$0.112 & 0.914$\pm$0.002 & 0.139$\pm$0.001 \\
    SM-CNN (Ours) & \textbf{31.359$\pm$0.119} & \underline{0.923$\pm$0.002} & \underline{0.124$\pm$0.001} \\
    \bottomrule
    \end{tabular}}
  \label{table:pu}
\end{table}

\begin{figure*}[t]
\centering
\includegraphics[width=5.4in]{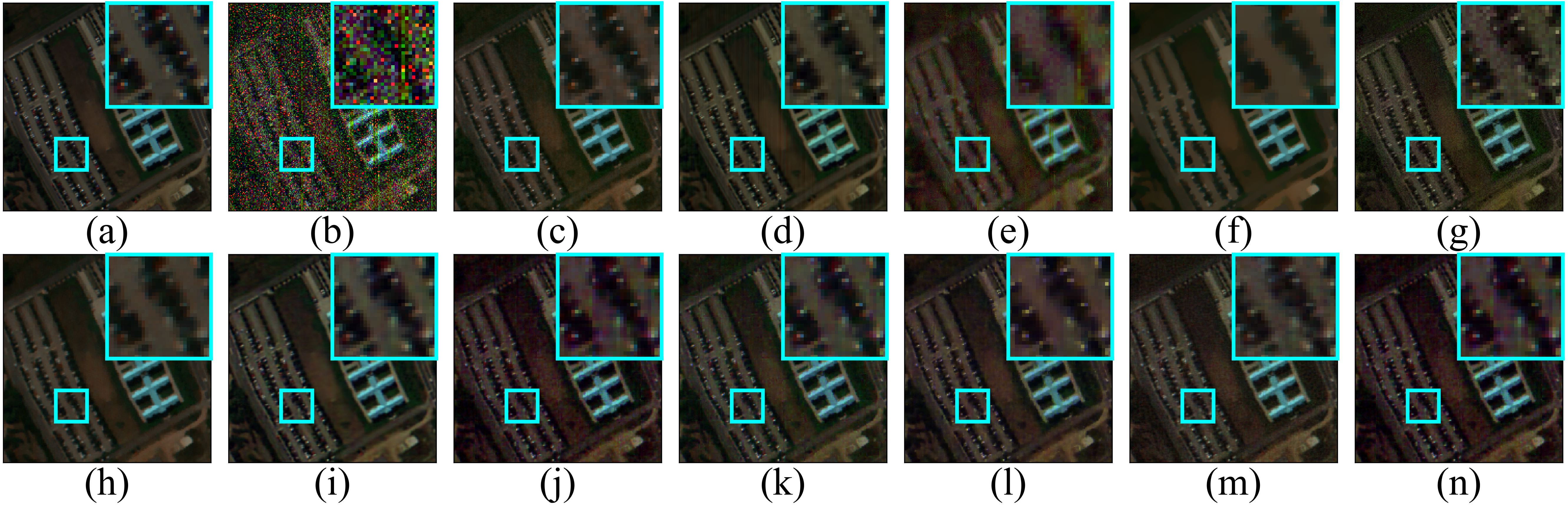}
\caption{Results for PU with mixture noise. (a) False-color original image with bands (60, 32, 10), (b) Noisy image, (c) LRTF-DFR, (d) FastHyMix, (e) BM4D, (f) LRTV, (g) LRMR, (h) LRTDTV, (i) QRNN3D, (j) HSID-CNN, (k) MemNet, (l) HDNET, (m) MAN, (n) SM-CNN (Ours).}
\label{fig_pu_mixture}
\end{figure*}

Table \ref{table:pu} shows the quantitative evaluation of the denoising results of different methods for the PU dataset distorted by mixture noise. \textcolor{black}{We ran tests on 10 different scenarios to observe the performance changes of all methods when different bands were corrupted with sparse noise. The table shows the mean and standard deviations of the different runs that we conducted.} The proposed SM-CNN achieves the highest MPSNR, second best MSSIM and SAM for the PU dataset without any fine tuning. QRNN3D achieves the highest MSSIM and SAM, which may be due to the possibility that the training data will better match these data. \textcolor{black}{Because MAN uses 3D convolution, retraining it when the number of bands in the data changes will improve its performance and adaptability. HSID-CNN, MemNet and HDNET} have performed worse than the other deep models, and even the classical LRTDTV method outperformed them in terms of metrics. 

\begin{figure*}[hbt!]
\centering
\includegraphics[width=4.0in]{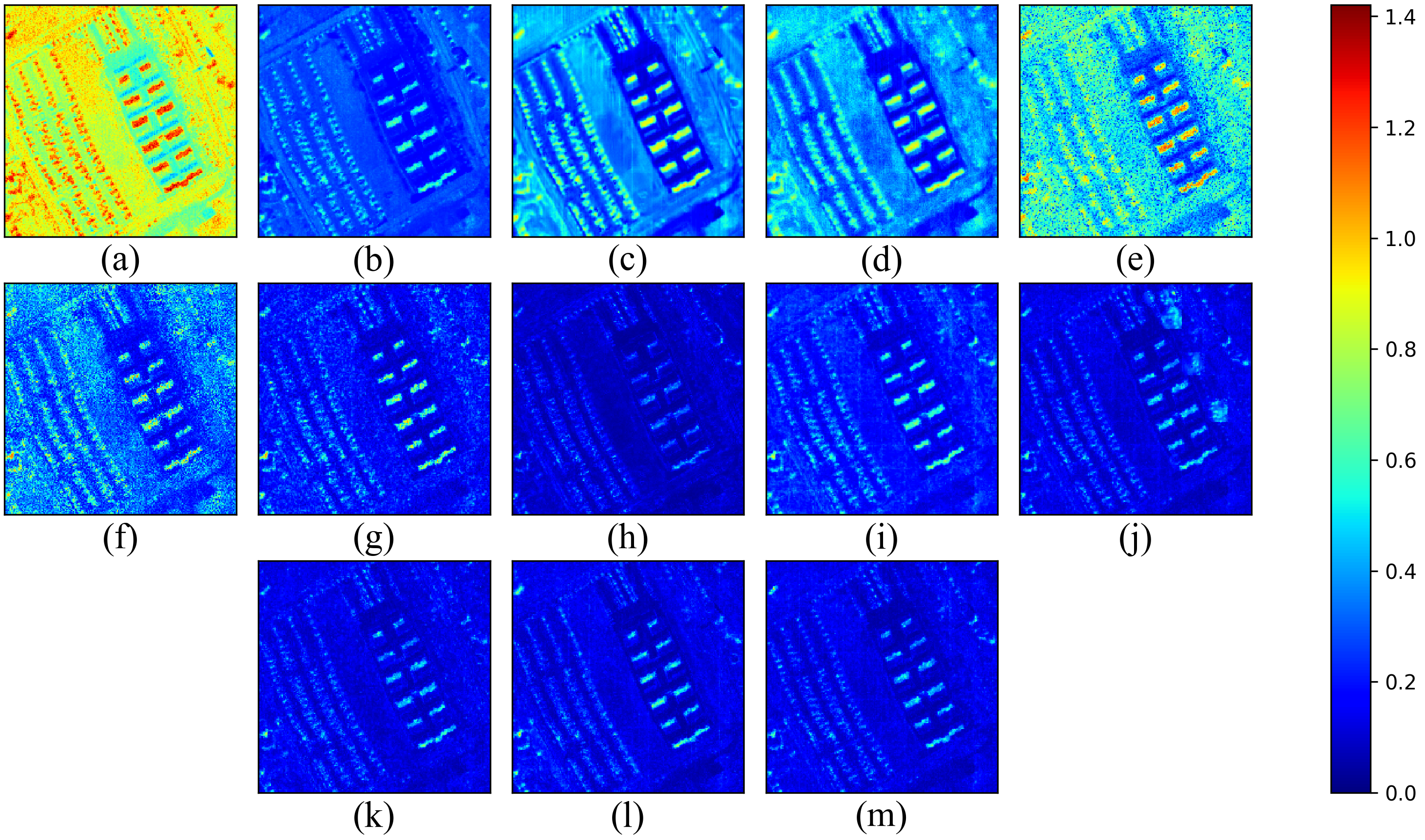}
\caption{SAM values corresponding to each pixel of the PU dataset: (a) Noisy data, (b) LRTF-DFR, (c) FastHyMix, (d) BM4D, (e) LRTV, (f) LRMR, (g) LRTDTV, (h) QRNN3D, (i) HSID-CNN, (j) MemNet, (k) HDNET, (l) MAN, (m) SM-CNN (Ours).}
\label{fig_pu_sam_all}
\end{figure*}

\begin{figure}[hbt!]
\centering
\includegraphics[width=3.5in]{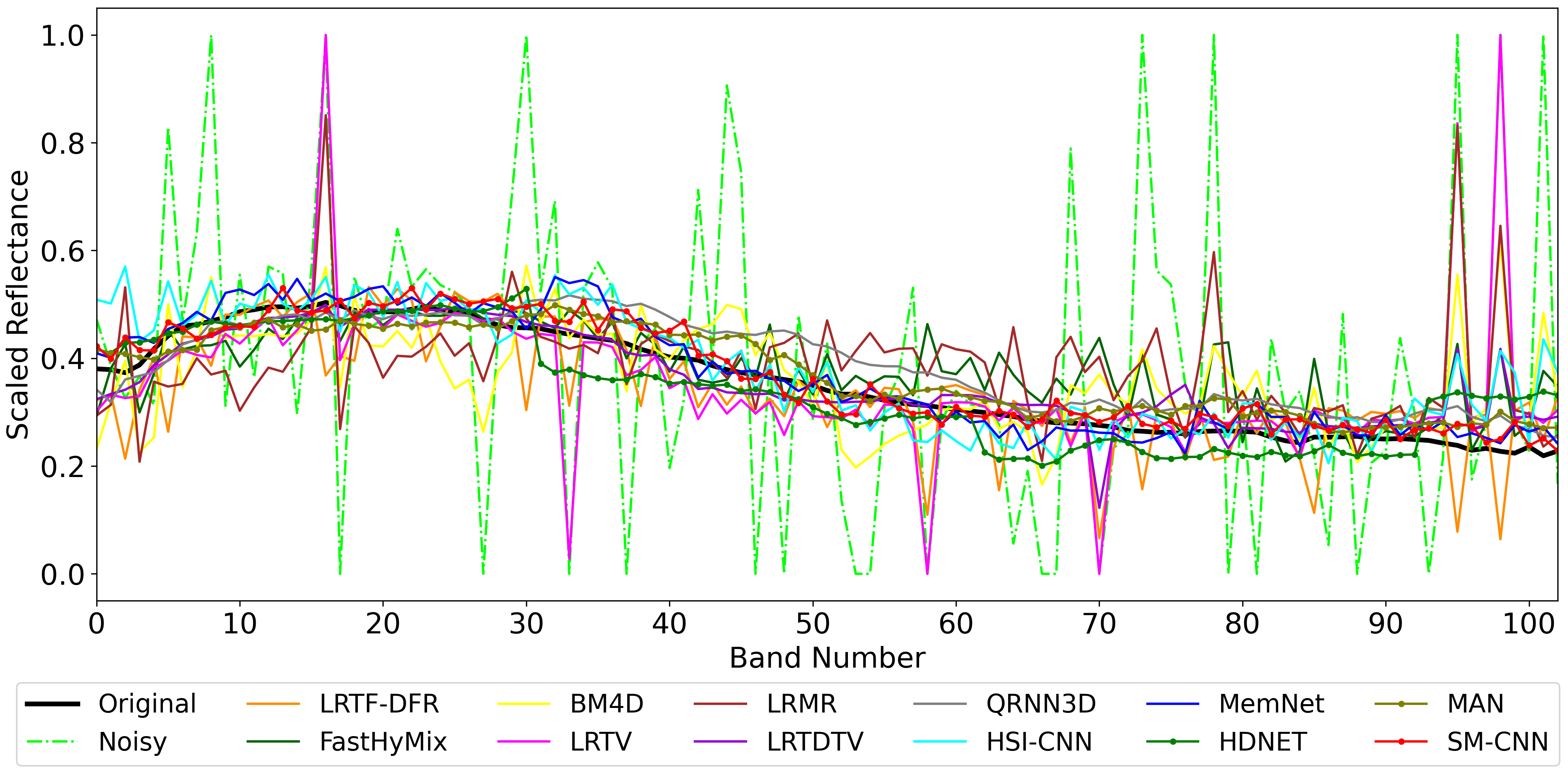}
\caption{Quality of spectral signature restoration of pixel (55, 169) in PU data.}
\label{fig_pu_signatures}
\end{figure}

\textcolor{black}{For qualitative evaluation, Fig. \ref{fig_pu_mixture} shows false-color RGBs of noisy data and denoising results for visual comparison. Looking at both visual results and quantitative metrics, we see that BM4D does not quite deal with mixture noise. In Fig. \ref{fig_pu_mixture} (e), it is clear that the SN is not removed. Additionally, BM4D result shows the effect of excessive smoothing as it destroys all the details in the zoomed area. The denoising performance of FastHyMix seems to be better than BM4D, but it slightly changes the color intensity of the image, which is evident from the roof area in Fig. \ref{fig_pu_mixture} (d). LRMR, LRTV and especially LRTDTV, which are more suitable for mixture noise removal, produce satisfactory outcomes in terms of both quantity and quality. Although not as much as BM4D, TV norm-based methods smooth the image as displayed in Fig. \ref{fig_pu_mixture} (f), (h). According to the result displayed in Fig. \ref{fig_pu_mixture} (c), it appears that LRTF-DFR is more effective at restoring details compared to traditional methods for these particular bands. Nevertheless, as indicated in Table \ref{table:pu}, the MPSNR and MSSIM metrics of LRTF-DFR are lower than those of LRTDTV, it can be argued that some bands may not have been corrected sufficiently. HSID-CNN seems to change the color intensity of the image for these bands. Our proposed model performs better in terms of preserving details while eliminating noise as shown in the zoomed regions in Fig. \ref{fig_pu_mixture}. Considering the details for these bands, \textcolor{black}{deep models} do not produce as good results as our model.}

Fig. \ref{fig_pu_sam_all} shows the SAM values corresponding to each pixel of the noisy PU dataset and the SAM values of denoising results of all methods in the dimensions of the original image. In other words, it shows how much the simulated mixture noise distorts each pixel spectrally, and the success of the denoising methods in restoring this distorted spectral information. The values given in Table \ref{table:pu} for the relevant method are obtained by averaging the SAM values in Fig. \ref{fig_pu_sam_all}. Among the classical methods, it is seen that the LRTDTV achieves the best results in most pixels. In addition, it is clear that DL methods give better results than classical methods. When we compare the DL methods, the performance of the methods decreases in the roof regions where the deterioration is intense. In particular, QRNN3D achieves better results than our SM-CNN in these regions. For this reason, it has obtained a slightly better result from our method on the average as seen from Table \ref{table:pu}.

In addition to the SAM metric, spectral signatures at (59,169) pixel of the PU data are given in Fig. \ref{fig_pu_signatures} to show the quality of the spectral signature restoration for each framework. When the noisy signature is examined, it can be seen that this pixel is generally distorted by zero-mean GN. In some bands, a value of one means that it is disturbed by SN or IN, while a value of zero can be said to be exposed to DN. Often times, classical methods fail to restore spectral information. For example, these methods, notably LRTV, are unable to recover signature from IN, SN and DN at some points. In DL approaches, HSID-CNN has not successfully recovered the signature in the first bands and last bands for this pixel, which was mainly distorted by IN or SN. We see that QRNN3D \textcolor{black}{and HDNET create} a bias in the middle bands for this pixel. MemNet, \textcolor{black}{MAN} and the proposed method produce a result very close to the original signature at every band. Furthermore, observing the results of the SAM metric, which is a measure of the spectral fidelity of all pixels, it can be said that the QRNN3D and proposed method achieve the best results.

\subsection{Real-Data Experiments}

\begin{figure*}[htb!]
\centering
\includegraphics[width=5.5in]{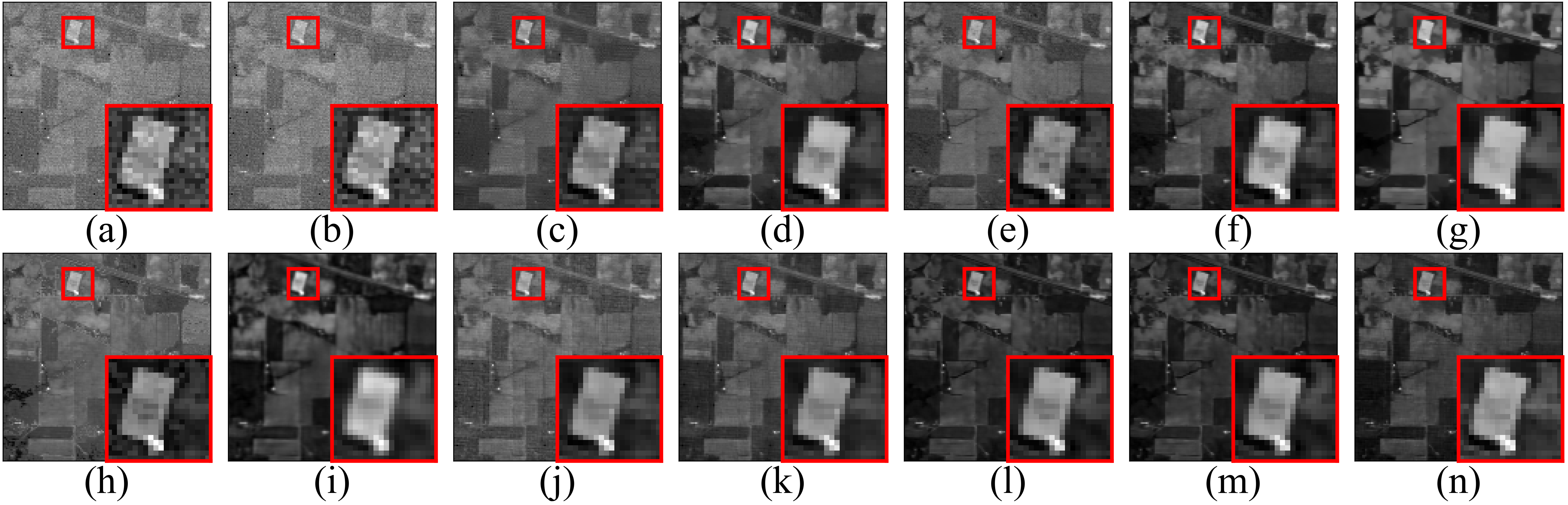}
\caption{Results on the IP dataset. (a) Grayscale visualization of noisy HSI using band 2, (b) TDL, (c) BM4D, (d) LRTV, (e) LRMR, (f) LRTDTV, (g) LRTF-DFR, (h) FastHyMix, (i) QRNN3D (j) HSID-CNN, (k) MemNet, (l) HDNET, (m) MAN, (n) SM-CNN (Ours).}
\label{fig_ip}
\end{figure*}

\noindent \textbf{IP dataset.} We also validate our model in the real-world noisy IP HSI without the corresponding ground truth. The first few bands and several other bands of IP the dataset are severely degraded by unknown noise. Since we do not have a reference clean image for IP, the performance of the methods have been evaluated via an SVM classifier. SVM has been trained with $10\%$  of randomly generated samples from each class. In the real experiment, 16 ground truth classes have been used to test the classification accuracy. Table \ref{table:ip} shows the classification results obtained after applying the denoising methods \textcolor{black}{and the run-times of each method.} The first row in the table shows the methods, while the second and third rows show the OA and kappa coefficient obtained using the denoising results, respectively. \textcolor{black}{We include the run-time information in the last row for reference only. We want to highlight that we did not engage in any optimization efforts on our code to enhance its run-time.} The HSID-CNN result is obtained using a pre-trained network with GN which has a variance equal to 50. The result of QRNN3D is obtained by the pre-trained network, which is first trained on GN with a variance equal to 50 followed by training on GN with variable variance. Because these networks cannot adapt their structure to the noise at the input like the one we propose, the noise used in the training and the test noise should match. On the other hand, no special training has been performed on MemNet, \textcolor{black}{HDNET, MAN} and the proposed SM-CNN for this dataset.

\renewcommand*{\arraystretch}{1.5}
\begin{table*}[hbt!]
\caption{Classification accuracies on the IP Dataset (SVM, $10\%$ training labels) {\textcolor{black} {together with a runtime analysis of the methods.}}}
    \centering
    \scalebox{0.50}{
    \begin{tabular}{p{0.10\textwidth}>{\centering}p{0.10\textwidth}>{\centering}p{0.10\textwidth}>{\centering}p{0.10\textwidth}>{\centering}p{0.10\textwidth}>{\centering}p{0.10\textwidth}>{\centering}p{0.10\textwidth}>{\centering}p{0.10\textwidth}>{\centering}p{0.10\textwidth}>{\centering}p{0.10\textwidth}>{\centering}p{0.10\textwidth}>{\centering}p{0.10\textwidth}>{\centering}p{0.10\textwidth}>{\centering}p{0.10\textwidth}>{\centering\arraybackslash}p{0.10\textwidth}}
    \toprule
    \multirow{2}{*}{Metrics} & Noisy & TDL & BM4D  & LRTV  & LRMR  & LRTD-TV  & LRTF-DFR & FastHy-Mix & QRNN-3D  & HSID-CNN  & MemNet  & \textcolor{black}{HDNET} & \textcolor{black}{MAN} & SM-CNN \\
     && \cite{peng2014TDL} &\cite{maggioni2012bm4d}&\cite{he2015total}&\cite{Zhang2014lrmr}&\cite{wang2017hyperspectral}&\cite{zheng2020double}&\cite{zhuang2021fasthymix}&\cite{wei20203}&\cite{yuan2019Hyperspectral}& \cite{tai2017memnet} & \cite{hu2022hdnet} &\cite{lai2023mixed} & (Ours)\\
    \midrule
    OA$\uparrow$ & 75.79 & 76.79 & 83.97 &78.72& 79.44 & 78.91& 88.49& 76.57& \underline{89.07}&86.65&88.70& 87.20 &87.33 &\textbf{89.31}\\
    Kappa$\uparrow$ & 0.7218  & 0.7305 & 0.8171  &0.7553 & 0.7579  &0.7641 & 0.8687 & 0.7306 &\underline{0.8747} & 0.8338& 0.8713& 0.8533& 0.8549&\textbf{0.8781}\\
    \midrule
    \multicolumn{2}{l}{\textcolor{black}{Run-Time (s)}}  & 15.295 & 285.151  &656.192 & 61.842  &229.315 &63.298  & 0.287& 13.090 &5.856 & 22.826& 6.145& 8.887&12.465\\
    \bottomrule
    \end{tabular}}
    
    \label{table:ip}
\end{table*}

In Fig. \ref{fig_ip}, the 2nd band of the IP dataset obtained by all methods is presented as a grayscale image for visual comparison. \textcolor{black}{According to the classification metrics, it can be seen that the performance of TDL and FastHyMix in real noise are not good enough. It is understood from metric results that BM4D gives better results. The complex noise removal methods LRMR, LRTV and LRTDTV seem to have good denoising performance for this band, but according to the classification results, it is clear that their performance is lower than BM4D. It can be observed that among the classical methods, LRTF-DFR yields the most favorable outcomes.} Among the DL methods, the QRNN3D method introduces blurring for this band as shown in Fig. \ref{fig_ip}(i). Our proposed method, on the other hand, produces visually sharper results than the other methods. Moroever, proposed SM-CNN shows higher performance than all methods according to the classification results given in Table \ref{table:ip}. The point that should not be forgotten here is that HSID-CNN and QRNN3D lag behind the proposed method even though they were trained with a special case to get the best results on the test data.

\begin{figure*}[htb!]
\centering
\includegraphics[width=5.5in]{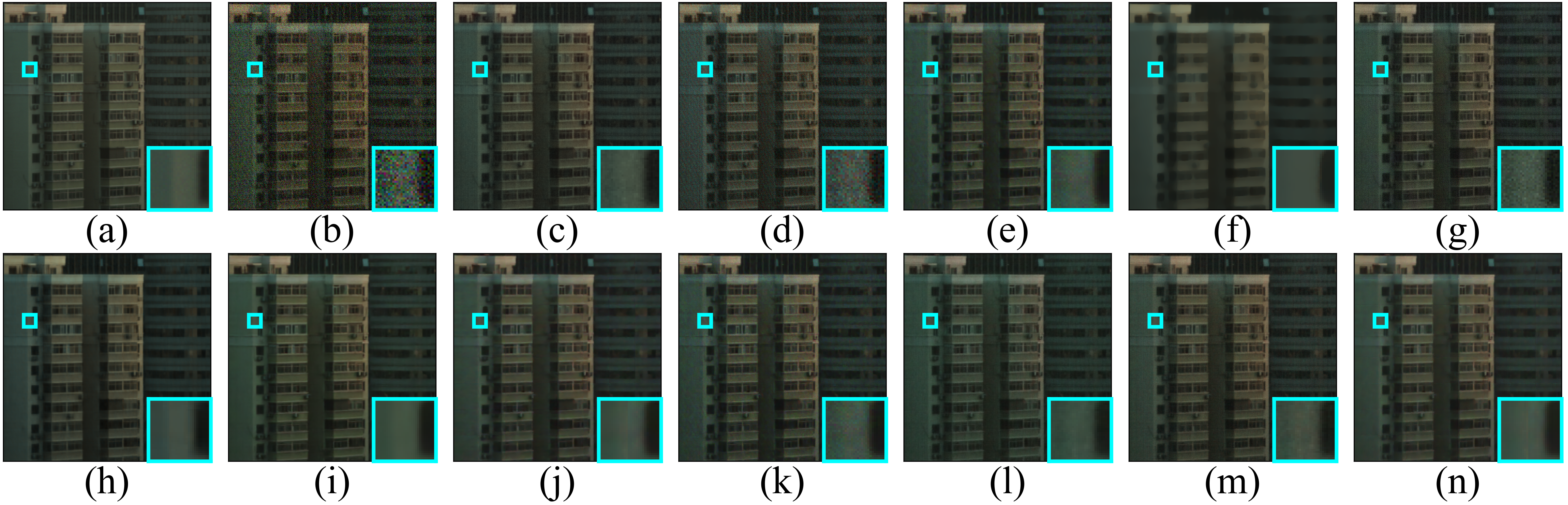}
\caption{Results for HSIDwRD with real noise. (a) Long-exposure false-color clean image with bands (30, 15, 10), (b) Short-exposure noisy image, (c) LRTF-DFR, (d) FastHyMix, (e) BM4D, (f) LRTV, (g) LRMR, (h) LRTDTV, (i) QRNN3D, (j) HSID-CNN, (k) MemNet, (l) HDNET, (m) MAN, (n) SM-CNN (Ours).}
\label{fig_HSIDwRD}
\end{figure*}

\renewcommand*{\arraystretch}{1.5}
\begin{table*}[hbt!]
\caption{Quantitative evaluation of different denoising methods on the HSIDwRD dataset.}
    \centering
    \scalebox{0.54}{
    \begin{tabular}{p{0.10\textwidth}>{\centering}p{0.10\textwidth}>{\centering}p{0.10\textwidth}>{\centering}p{0.10\textwidth}>{\centering}p{0.10\textwidth}>{\centering}p{0.10\textwidth}>{\centering}p{0.10\textwidth}>{\centering}p{0.10\textwidth}>{\centering}p{0.10\textwidth}>{\centering}p{0.10\textwidth}>{\centering}p{0.10\textwidth}>{\centering}p{0.10\textwidth}>{\centering}p{0.10\textwidth}>{\centering\arraybackslash}p{0.10\textwidth}}
    \toprule
    \multirow{2}{*}{Metrics} & Noisy & LRTF-DFR & FastHy-Mix & BM4D  & LRTV  & LRMR  & LRTD-TV  & QRNN-3D  & HSID-CNN  & MemNet  & \textcolor{black}{HDNET} & \textcolor{black}{MAN} &SM-CNN \\
     &&\cite{zheng2020double}&\cite{zhuang2021fasthymix}&\cite{maggioni2012bm4d}&\cite{he2015total}&\cite{Zhang2014lrmr}&\cite{wang2017hyperspectral}&\cite{wei20203}&\cite{yuan2019Hyperspectral}& \cite{tai2017memnet} & \cite{hu2022hdnet} &\cite{lai2023mixed} & (Ours)\\
    \midrule
    MPSNR$\uparrow$ & 20.912 & 31.069 & 25.686 &30.573& 29.139 & 29.864& 29.599& \underline{31.894}&31.405&28.272& 30.801 &31.246 &\textbf{32.039}\\
    MSSIM$\uparrow$ & 0.358  & 0.922 & 0.649  &0.907 & 0.904  & 0.867 & 0.913 &  \underline{0.938} & 0.933& 0.890& 0.922& 0.923&\textbf{0.940}\\
    SAM$\downarrow$   & 0.552  & 0.147 & 0.335  &0.163 & 0.150  &0.185 & 0.154  &  \underline{0.140} &0.150 & 0.240& 0.147& 0.152&\textbf{0.139}\\
    \bottomrule
    \end{tabular}}
    
    \label{table:HSIDwRD}
\end{table*}

\noindent \textbf{HSIDwRD dataset.} \textcolor{black}{Real-Data experiment was also carried out on the real natural HSIDwRD dataset. The test was conducted using the WDC data-trained network without any fine-tuning. The quantitative outcomes are presented in Table \ref{table:HSIDwRD} and Fig. \ref{fig_HSIDwRD} shows long-exposure clean image, short-exposure noisy image and denoising results for visual comparison. The effectiveness of our method is evidenced by the fact that our results surpass the others in all metrics. In Fig. \ref{fig_HSIDwRD}, it can be seen that our method produces more sharp and clear results visually.}

\section{Ablation Study}\label{sec:5}
To verify the functionality of SSMRB component in our SM-CNN, some ablation experiments are conducted on the mixture noisy WDC. For this purpose, we present two more models, namely Wavelength Modulating
CNN (WM-CNN) and SM-CNN-Lite.

\noindent \textbf{WM-CNN:} As we mentioned in Sec.~\ref{sec:4-a}, some bands are omitted in HSI data due to atmospheric effects and other reasons. The proposed training uses information of neighboring spectral bands while eliminating the noise of one band; and due to these missing bands, the information of neighboring bands can change rapidly. Therefore, these missing bands affect the performance of the methods as seen in Fig. \ref{fig_psnr_ssim} around 0.9 $\mu$m. In addition, since HSI sensors collect data in different parts of the spectrum, the denoiser must also adapt to the wavelength information of different dataset. Considering these reasons, instead of the $K$ spectral bands in the SSMM module of the denoiser, the wavelength information of the denoising band is used. To do this, each element of a tensor with the same dimensions as the noisy patch to be denoised is placed in the value of the wavelength (in $\mu$m) of this band. To illustrate, for the noisy band collected at 0.4 $\mu$m wavelength, the modulation data in the SSMM is a tensor with size of 20$\times$20 and each value of this tensor is 0.4. Since this tensor is used as modulation data in the deep layers through SSMRB, we call it a Wavelength Modulating CNN. Finally, when the input size of the SSMM modules decrease from 20$\times$20$\times K$ to 20$\times$20, the number of trainable parameters of the WM-CNN is also less than our original~SM-CNN.

\noindent \textbf{SM-CNN-Lite:} In this model, we reduced the number of consecutive SSMRB modules in the SM-CNN shown in Fig. \ref{fig_network}. We use a single SSMRB instead of 2 SSMRB in the deep layer, but we add a layer with a kernel size of 1 in parallel with the convolution layer with the kernel size of 5 in the SSMM shown in Fig. \ref{fig_ain} in order to use spectral information directly. Then, we generate scale ($\mathbf{\gamma}$) and shift ($\mathbf{\beta}$) parameters by combining the results of these two layers. We call it SM-CNN-Lite when it uses the same modulation data as the original SM-CNN with less trainable parameters.

\renewcommand*{\arraystretch}{1.00}
\begin{table}[hbt!]
  \centering
  \caption{Quantitative evaluation on the WDC with mixture noise.}
    \scalebox{0.6}{\begin{tabular}{p{0.22\textwidth}>{\centering}p{0.12\textwidth}>{\centering}p{0.12\textwidth}>{\centering\arraybackslash}p{0.12\textwidth}>{\centering\arraybackslash}p{0.12\textwidth}}
    \toprule
    Method & MPSNR$\uparrow$ & MSSIM$\uparrow$ & SAM$\downarrow$ & NOP\\
    \midrule
    SM-CNN & \textbf{29.832} & \textbf{0.973} & \textbf{0.066} & 2,404,361\\
	WM-CNN    & 27.464 & 0.954 & 0.086 & \textbf{1,852,361} \\
	SM-CNN-Lite   & 28.570 & 0.964 & 0.075 & 1,867,241\\
    \bottomrule
    \end{tabular}}
  \label{table:ablation}
\end{table}

Table \ref{table:ablation} shows the quantitative evaluation of the denoising results of the evaluated versions of our model for the WDC dataset distorted by mixture noise and the number of parameter (NOP) of each model. Looking at the quantitative results, both WM-CNN and SM-CNN-Lite yield the most favorable outcomes for the heavily distorted WDC dataset with different types of noise. That said, the best performance is achieved with our proposed model.

\begin{figure}[htb!]
\centering
\includegraphics[width=3.5in]{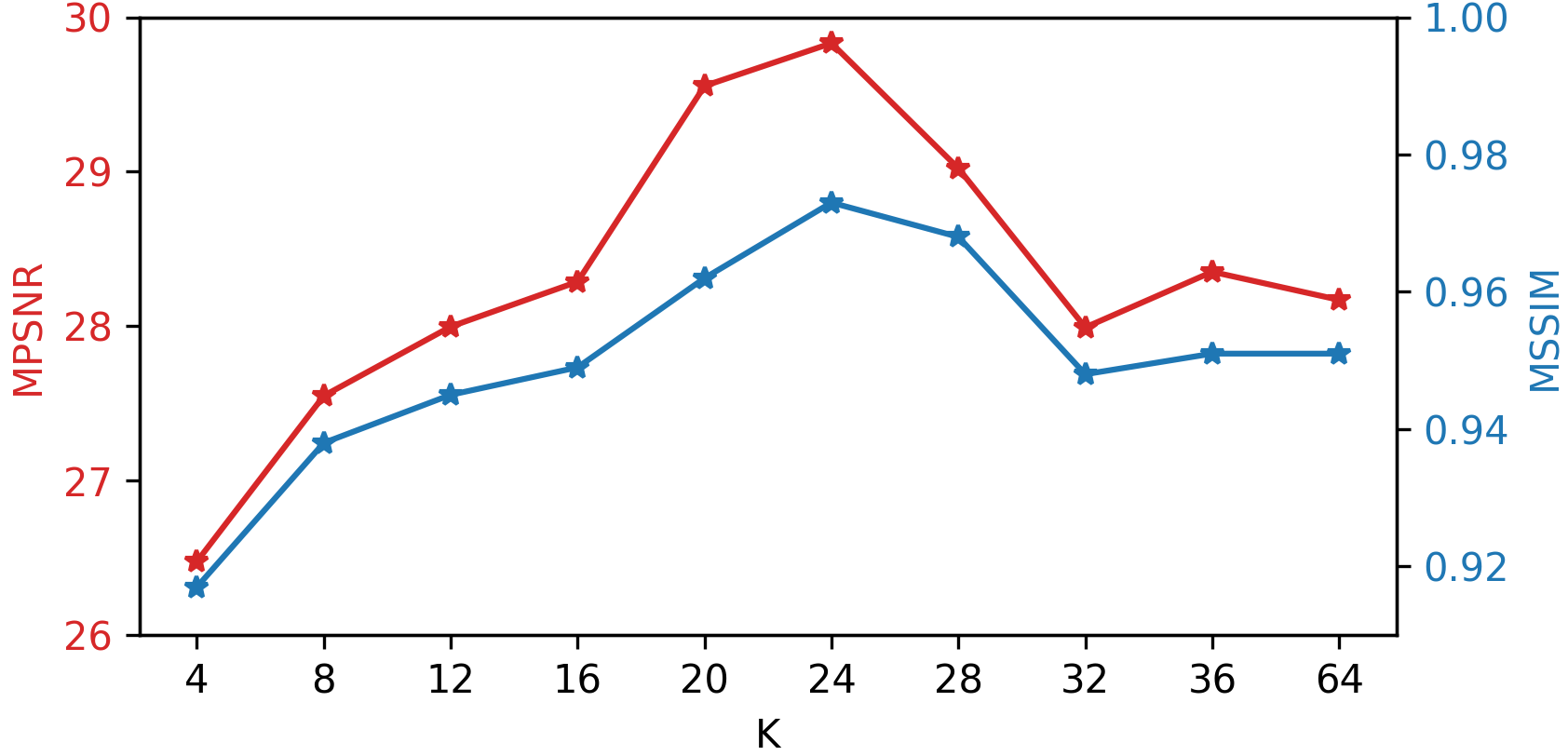}
\caption{Effectiveness of denoising with different adjacent spectral band.}
\label{fig_k_psnr_ssim}
\quad
\end{figure}

\noindent \textcolor{black}{\textbf{Investigating the Impact of Various $K$ values:}} \textcolor{black}{In Fig. \ref{fig_k_psnr_ssim}, we explore the influence of varying the number of neighboring bands ($K$) on our network's performance. The network has been trained using 10 different values of $K$ on the WDC, with mixture noise. As illustrated in Fig. \ref{fig_k_psnr_ssim}, employing a limited number of bands leads to poor results, revealing the significance of considering a broader spectral context. As the number of bands increases, we witness an improvement in performance; but it does not have a linear trend. Instead, it exhibits slight fluctuations as the number of neighboring bands increases, unveiling subtle changes in the signals' spectral characteristics influenced by neighboring bands.}

\noindent \textcolor{black}{\textbf{Investigating the Impact of the Number of Skip Connections:}} \textcolor{black}{Table \ref{table:sc} presents a comprehensive investigation into the significance of skip connections within the proposed SM-CNN framework. Through this systematic analysis, conducted on the WDC dataset with the mixture noise scenario, we gradually remove skip connections, beginning from the network's input layer, and perform training accordingly. The results clearly demonstrate the pivotal role played by skip connections in enhancing the network's performance. Remarkably, as the number of skip connections decreases, we observe a substantial decline in the denoising performance of our method. The incorporation of skip connections ensures better convergence during training, promoting effective information flow and preserving essential features within the denoising process.}
\renewcommand*{\arraystretch}{1.00}
\begin{table}[hbt!]
  \centering
  \caption{Impact of skip connections on network performance with K=24.}
    \scalebox{0.6}{\begin{tabular}{p{0.22\textwidth}>{\centering}p{0.12\textwidth}>{\centering}p{0.12\textwidth}>{\centering\arraybackslash}p{0.12\textwidth}>{\centering\arraybackslash}p{0.12\textwidth}}
    \toprule
    Metrics & Only last SC & Last two SC & Last three SC & All SC\\
    \midrule
    MPSNR & 28.766 & 28.806 & 29.186 & \textbf{29.832}\\
	MSSIM    & 0.954 & 0.955 & 0.958   & \textbf{0.973} \\
    \bottomrule
    \end{tabular}}
  \label{table:sc}
\end{table}

\section{Conclusion}\label{sec:6}
In this work, we present an SM-CNN for complex noise reduction
in HSIs considering the noise type of GN, SN, IN, DN, mixture noise, and the real-world unknown noise. Thanks to the SM-CNN architecture, test data with different spatial-spectral properties can be denoised with a single model. \textcolor{black}{Moreover, modulating the network using spatio-spectral information through the SSMRB enables its adaptation to different noise by integrating input data-driven dynamic predicted features.} The qualitative and quantitative evaluation of the results show that the proposed algorithm is more efficient than other single-model algorithms on both synthetic and real data.

\section{Acknowledgments}
O. Torun’s PhD research has been partially supported by the KUIS AI Research Center and the 2023 BAGEP Award, which was granted to S. E. Yuksel by the Science Academy.

 \bibliographystyle{elsarticle-num} 
 \bibliography{ref}

\begin{thebibliography}{10}
\expandafter\ifx\csname url\endcsname\relax
  \def\url#1{\texttt{#1}}\fi
\expandafter\ifx\csname urlprefix\endcsname\relax\def\urlprefix{URL }\fi
\expandafter\ifx\csname href\endcsname\relax
  \def\href#1#2{#2} \def\path#1{#1}\fi

\bibitem{kucuk2021total}
S.~Kucuk, S.~E. Yuksel, Total utility metric based dictionary pruning for sparse hyperspectral unmixing, IEEE Trans. Comput. Imaging 7 (2021) 562--572.

\bibitem{qi2020deep}
L.~Qi, J.~Li, Y.~Wang, M.~Lei, X.~Gao, Deep spectral convolution network for hyperspectral image unmixing with spectral library, Signal Process. 176 (2020) 107672.

\bibitem{lu2023ensemble}
Y.~Lu, X.~Zheng, H.~Xin, H.~Tang, R.~Wang, F.~Nie, Ensemble and random collaborative representation-based anomaly detector for hyperspectral imagery, Signal Process. 204 (2023) 108835.

\bibitem{rasti2018noise}
B.~Rasti, P.~Scheunders, P.~Ghamisi, G.~Licciardi, J.~Chanussot, Noise reduction in hyperspectral imagery: Overview and application, Remote Sensing 10~(3) (2018) 482.

\bibitem{zhang2019Hybrid}
Q.~Zhang, Q.~Yuan, J.~Li, X.~Liu, H.~Shen, L.~Zhang, Hybrid noise removal in hyperspectral imagery with a spatial--spectral gradient network, IEEE Trans. Geosci. Remote Sens. 57~(10) (2019) 7317--7329.

\bibitem{bahraini2022bayesian}
T.~Bahraini, A.~Ebrahimi-Moghadam, M.~Khademi, H.~S. Yazdi, Bayesian framework selection for hyperspectral image denoising, Signal Process. 201 (2022) 108712.

\bibitem{zhang2021hyperspectral}
T.~Zhang, Y.~Fu, C.~Li, Hyperspectral image denoising with realistic data, in: IEEE/CVF ICCV, 2021, pp. 2248--2257.

\bibitem{yuan2019Hyperspectral}
Q.~Yuan, Q.~Zhang, J.~Li, H.~Shen, L.~Zhang, Hyperspectral image denoising employing a spatial–spectral deep residual convolutional neural network, IEEE Trans. Geosci. Remote Sens. 57~(2) (2019) 1205--1218.

\bibitem{maffei2020single}
A.~Maffei, J.~M. Haut, M.~E. Paoletti, J.~Plaza, L.~Bruzzone, A.~Plaza, A single model cnn for hyperspectral image denoising, IEEE Trans. Geosci. Remote Sens. 58~(4) (2020) 2516--2529.

\bibitem{wei20203}
K.~Wei, Y.~Fu, H.~Huang, 3-d quasi-recurrent neural network for hyperspectral image denoising, IEEE Trans. Neural Netw. Learn Syst. 32~(1) (2021) 363--375.

\bibitem{wang2022sscan}
Z.~Wang, Z.~Shao, X.~Huang, J.~Wang, T.~Lu, Sscan: A spatial–spectral cross attention network for hyperspectral image denoising, IEEE Geosci. Remote. Sens. Lett. 19 (2022) 1--5.

\bibitem{han2021dynamic}
Y.~Han, G.~Huang, S.~Song, L.~Yang, H.~Wang, Y.~Wang, Dynamic neural networks: A survey, IEEE Trans. Pattern Anal. Mach. Intell. 44~(11) (2021) 7436--7456.

\bibitem{li2023spectral}
M.~Li, J.~Liu, Y.~Fu, Y.~Zhang, D.~Dou, Spectral enhanced rectangle transformer for hyperspectral image denoising, in: Proceedings of the IEEE/CVF CVPR, 2023, pp. 5805--5814.

\bibitem{Zhang2014lrmr}
H.~Zhang, W.~He, L.~Zhang, H.~Shen, Q.~Yuan, Hyperspectral image restoration using low-rank matrix recovery, IEEE Trans. Geosci. Remote Sens. 52~(8) (2014) 4729--4743.

\bibitem{he2015hyperspectral}
W.~He, H.~Zhang, L.~Zhang, H.~Shen, Hyperspectral image denoising via noise-adjusted iterative low-rank matrix approximation, IEEE J. Sel. Top. Appl. Earth Obs. Remote Sens. 8~(6) (2015) 3050--3061.

\bibitem{he2015total}
W.~He, H.~Zhang, L.~Zhang, H.~Shen, Total-variation-regularized low-rank matrix factorization for hyperspectral image restoration, IEEE Trans. Geosci. Remote Sens. 54~(1) (2016) 178--188.

\bibitem{wang2017hyperspectral}
Y.~Wang, J.~Peng, Q.~Zhao, Y.~Leung, X.-L. Zhao, D.~Meng, Hyperspectral image restoration via total variation regularized low-rank tensor decomposition, IEEE J. Sel. Top. Appl. Earth Obs. Remote Sens. 11~(4) (2018) 1227--1243.

\bibitem{zhang20233d}
F.~Zhang, K.~Zhang, W.~Wan, J.~Sun, 3d geometrical total variation regularized low-rank matrix factorization for hyperspectral image denoising, Signal Process. (2023) 108942.

\bibitem{yuan2012hyperspectral}
Q.~Yuan, L.~Zhang, H.~Shen, Hyperspectral image denoising employing a spectral--spatial adaptive total variation model, IEEE Trans. Geosci. Remote Sens. 50~(10) (2012) 3660--3677.

\bibitem{zheng2020double}
Y.-B. Zheng, T.-Z. Huang, X.-L. Zhao, Y.~Chen, W.~He, Double-factor-regularized low-rank tensor factorization for mixed noise removal in hyperspectral image, IEEE Trans. Geosci. Remote Sens. 58~(12) (2020) 8450--8464.

\bibitem{he2021tslrln}
C.~He, L.~Sun, W.~Huang, J.~Zhang, Y.~Zheng, B.~Jeon, Tslrln: Tensor subspace low-rank learning with non-local prior for hyperspectral image mixed denoising, Signal Process. 184 (2021) 108060.

\bibitem{zeng2021hyperspectral}
H.~Zeng, X.~Xie, J.~Ning, Hyperspectral image denoising via global spatial-spectral total variation regularized nonconvex local low-rank tensor approximation, Signal Process. 178 (2021) 107805.

\bibitem{liu2022multi}
N.~Liu, W.~Li, R.~Tao, Q.~Du, J.~Chanussot, Multi-graph-based low-rank tensor approximation for hyperspectral image restoration, IEEE Trans. Geosci. Remote Sens. 60 (2022) 1--14.

\bibitem{zhang2022double}
H.~Zhang, J.~Cai, W.~He, H.~Shen, L.~Zhang, Double low-rank matrix decomposition for hyperspectral image denoising and destriping, IEEE Trans. Geosci. Remote Sens. 60 (2022) 1--19.

\bibitem{chen2022hyperspectral}
Y.~Chen, W.~Cao, L.~Pang, X.~Cao, Hyperspectral image denoising with weighted nonlocal low-rank model and adaptive total variation regularization, IEEE Trans. Geosci. Remote Sens. (2022) 1--1.

\bibitem{liu2023survey}
N.~Liu, W.~Li, Y.~Wang, R.~Tao, Q.~Du, J.~Chanussot, A survey on hyperspectral image restoration: From the view of low-rank tensor approximation, Sci. China Inf. Sci. 66~(4) (2023) 140302.

\bibitem{peng2022low}
J.~Peng, W.~Sun, H.-C. Li, W.~Li, X.~Meng, C.~Ge, Q.~Du, Low-rank and sparse representation for hyperspectral image processing: A review, IEEE Geosci. Remote Sens. Mag. 10~(1) (2022) 10--43.

\bibitem{maggioni2012bm4d}
M.~Maggioni, V.~Katkovnik, K.~Egiazarian, A.~Foi, Nonlocal transform-domain filter for volumetric data denoising and reconstruction, IEEE Trans. Image Process. 22~(1) (2013) 119--133.

\bibitem{Dabov2007bm3d}
K.~Dabov, A.~Foi, V.~Katkovnik, K.~Egiazarian, Image denoising by sparse 3-d transform-domain collaborative filtering, IEEE Trans. Image Process. 16~(8) (2007) 2080--2095.

\bibitem{he2019non}
W.~He, Q.~Yao, C.~Li, N.~Yokoya, Q.~Zhao, H.~Zhang, L.~Zhang, Non-local meets global: An iterative paradigm for hyperspectral image restoration, IEEE Trans. Pattern Anal. Mach. Intell. 44~(4) (2022) 2089--2107.

\bibitem{zhuang2018fast}
L.~Zhuang, J.~M. Bioucas-Dias, Fast hyperspectral image denoising and inpainting based on low-rank and sparse representations, IEEE J. Sel. Top. Appl. Earth Obs. Remote Sens. 11~(3) (2018) 730--742.

\bibitem{zhuang2021fasthymix}
L.~Zhuang, M.~K. Ng, Fasthymix: Fast and parameter-free hyperspectral image mixed noise removal, IEEE Trans Neural Netw Learn Syst (2021) 1--15.

\bibitem{zhao2022hyperspectral}
B.~Zhao, M.~O. Ulfarsson, J.~R. Sveinsson, J.~Chanussot, Hyperspectral image denoising using spectral-spatial transform-based sparse and low-rank representations, IEEE Trans. Geosci. Remote Sens. 60 (2022) 1--25.

\bibitem{zhang2017beyond}
K.~Zhang, W.~Zuo, Y.~Chen, D.~Meng, L.~Zhang, Beyond a gaussian denoiser: Residual learning of deep cnn for image denoising, IEEE Trans. Image Process. 26~(7) (2017) 3142--3155.

\bibitem{chang2018hsi}
Y.~Chang, L.~Yan, H.~Fang, S.~Zhong, W.~Liao, Hsi-denet: Hyperspectral image restoration via convolutional neural network, IEEE Trans. Geosci. Remote Sens. 57~(2) (2019) 667--682.

\bibitem{tai2017memnet}
Y.~Tai, J.~Yang, X.~Liu, C.~Xu, Memnet: A persistent memory network for image restoration, in: IEEE ICCV, 2017, pp. 4539--4547.

\bibitem{sidorov2019deep}
O.~Sidorov, J.~Yngve~Hardeberg, Deep hyperspectral prior: Single-image denoising, inpainting, super-resolution, in: IEEE/CVF ICCVW, 2019.

\bibitem{miao2022hyperspectral}
Y.-C. Miao, X.-L. Zhao, X.~Fu, J.-L. Wang, Y.-B. Zheng, Hyperspectral denoising using unsupervised disentangled spatiospectral deep priors, IEEE Trans. Geosci. Remote Sens. 60 (2022) 1--16.

\bibitem{pan2022sqad}
E.~Pan, Y.~Ma, X.~Mei, F.~Fan, J.~Huang, J.~Ma, Sqad: Spatial-spectral quasi-attention recurrent network for hyperspectral image denoising, IEEE Trans. Geosci. Remote Sens. 60 (2022) 1--14.

\bibitem{wang2022translution}
C.~Wang, M.~Xu, Y.~Jiang, et~al., Translution-snet: A semisupervised hyperspectral image stripe noise removal based on transformer and cnn, IEEE Trans. Geosci. Remote Sens. 60 (2022) 1--14.

\bibitem{xiong2022mac}
F.~Xiong, J.~Zhou, Q.~Zhao, J.~Lu, Y.~Qian, Mac-net: Model-aided nonlocal neural network for hyperspectral image denoising, IEEE Trans. Geosci. Remote Sens. 60 (2022) 1--14.

\bibitem{lai2023mixed}
Z.~Lai, Y.~Fu, Mixed attention network for hyperspectral image denoising, arXiv preprint arXiv:2301.11525 (2023).

\bibitem{hu2022hdnet}
X.~Hu, Y.~Cai, J.~Lin, H.~Wang, X.~Yuan, Y.~Zhang, R.~Timofte, L.~Van~Gool, Hdnet: High-resolution dual-domain learning for spectral compressive imaging, in: Proceedings of the IEEE/CVF CVPR, 2022, pp. 17542--17551.

\bibitem{ronneberger2015u}
O.~Ronneberger, P.~Fischer, T.~Brox, U-net: Convolutional networks for biomedical image segmentation, in: International Conference on MICCAI, Springer, 2015, pp. 234--241.

\bibitem{peng2014TDL}
Y.~Peng, D.~Meng, Z.~Xu, C.~Gao, Y.~Yang, B.~Zhang, Decomposable nonlocal tensor dictionary learning for multispectral image denoising, in: IEEE CVPR, 2014, pp. 2949--2956.

\bibitem{archibald2007feature}
R.~Archibald, G.~Fann, Feature selection and classification of hyperspectral images with support vector machines, IEEE Geosci. Remote Sens. Lett. 4~(4) (2007) 674--677.

\bibitem{wang2014wdc}
Q.~Wang, L.~Zhang, Q.~Tong, F.~Zhang, Hyperspectral imagery denoising based on oblique subspace projection, IEEE J. Sel. Top. Appl. Earth Obs. Remote Sens. 7~(6) (2014) 2468--2480.

\bibitem{baumgardner2015220}
M.~F. Baumgardner, L.~L. Biehl, D.~A. Landgrebe, 220 band aviris hyperspectral image data set: June 12, 1992 indian pine test site 3, Purdue University Research Repository 10 (2015) R7RX991C.

\end{thebibliography}

\end{document}